\documentclass[preprint,authoryear,12pt]{elsarticle}

\usepackage{epsfig}

\usepackage{amssymb}

\usepackage{hyperref}

\makeatletter
\def\ps@pprintTitle{%
  \let\@oddhead\@empty
  \let\@evenhead\@empty
  \let\@oddfoot\@empty
  \let\@evenfoot\@oddfoot
}
\makeatother

\begin{document}

\begin{frontmatter}



\title{Radar-based Re-Entry Predictions with very limited tracking capabilities: the GOCE case study}


\author{S. Cical\`o \corref{cor}}
\address{Space Dynamics Services S.r.l., via Mario Giuntini 63, 56023 Navacchio (PI), Italy}
\cortext[cor]{Corresponding author}
\ead{cicalo@spacedys.com}


\author{S. Lemmens}
\address{ESA/ESOC Space Debris Office (OPS-GR), Robert-Bosch-Str. 5, 64293 Darmstadt, Germany}
\ead{Stijn.Lemmens@esa.int}


\begin{abstract}
  The problem of the re-entry predictions of GOCE has been
  investigated in many different ways and presented in the literature,
  especially because of the large amount of observational data, mainly
  radar and GPS, available until the very last part of its decay. The
  accurate GPS and attitude measurements can be used to compute a POD
  during the three weeks of decay, which can be exploited to
  extrapolate the ballistic coefficient evolution of the object, and
  as a reference trajectory. In previous works the main focus was on
  the german TIRA radar and on similar tracking sensors, to
  investigate the capabilities of radar-based orbit determination and
  ballistic coefficient calibration for re-entry predictions (residual
  lifetime estimation).

  During this work we have carried out additional analysis on
  radar-based re-entry predictions for GOCE, and for other similar
  decaying objects on circular and highly inclinated orbits. We focus
  on the northern european sensor EISCAT UHF radar, located in
  Troms\o, Norway. This sensor, originally conceived for atmospheric
  studies of the ionosphere, has been recently considered for space
  debris applications, for tracking of specific targets, and to
  support re-entry predictions. The limited tracking capabilities of
  this sensor poses the problem of establishing to which extent it
  would be useful to support orbit determination and re-entry
  predictions, in comparison to what we know about TIRA-like
  performances.

  We have set up a realistic simulation scenario for a re-entry
  campaign. Re-entry predictions and ballistic coefficient
  calibrations are performed and compared for TIRA-only, EISCAT-only,
  and both radar available situations. The results are compared in
  terms of differences in the orbital states over the total
  observation time span, in the ballistic coefficient estimation, and
  in the corresponding re-entry epoch.

  The main conclusion is that, provided a minimum amount of necessary
  observational information, EISCAT-based re-entry predictions are of
  comparable accuracy to TIRA-based (but also to GPS-based, and
  TLE-based if TLE errors are properly accounted) corresponding
  ones. Even if the worse tracking capabilities of the EISCAT sensor
  are not able to determine an orbit at the same level of accuracy of
  TIRA, it turned out that the estimated orbits are anyway
  equivalent in terms of re-entry predictions, if we consider the
  relevant parameters involved and their effects on the re-entry
  time. What happens to be very important is the difficulty in
  predicting both atmospheric and attitude significant variations in
  between the current epoch of observation and the actual
  re-entry. For this reason, it is not easy to keep the actual accuracy
  of the predictions much lower than 10\% of the residual lifetime,
  apart from particular cases with constant area to mass ratio, and
  low atmospheric environment variations with respect to current
  models.

  A corresponding GOCE ballistic coefficient estimation based on TLE
  only is presented, and it turns out to be very effective as
  well. Some critical cases which consist in a minimum amount of
  observational information, or in difficulties in obtaining OD
  convergence, are presented, and a list of possible countermeasures
  is proposed.

  An experiment with real EISCAT,TIRA, and TLE data is also
  presented, for the case of 2012-006K AVUM R/B. Additional
  experiments with simulated trajectories are presented as well, with
  analogous results.

\end{abstract}


\end{frontmatter}

\parindent=0.5 cm

\section{Introduction}
\label{sec:intro}

During the ESA GSP ``EXPRO+-Benchmarking Re-Entry Prediction
Uncertainties'' project, we have investigated the problem of the
estimation of the GOCE re-entry location, i.e. the re-entry time
prediction, on the basis of both continuous GPS measurements and
sparse radar tracking information (see \cite{cetal17}). The main
result is that, with a reasonable frequency of measurements,
radar-based Orbit Determination (OD) is very effective in estimating
the average evolution of the spacecraft's ballistic coefficient, in
comparison to the one estimated from continuous GPS measurements, even
from a single site. Focus was made on the german TIRA radar of the
Fraunhofer institute for High Frequency Physics and Radar Techniques,
and on analogous, even though less accurate, sensors. The high orbit
accuracy provided by GPS-based Precise Orbit Determination (POD) hides
the intrinsic large errors in the dynamical models, atmospheric
environment, and attitude behaviour, which are artificially absorbed
by the fitted empirical accelerations. These errors re-appear in the
radar-based OD under the form of large observational residuals. Even
with a very good knowledge of the current position and velocity of the
spacecraft, and a good average drag determination, in general it is
not possible to predict the future re-entry location with very high
accuracy (always much better than 10\%). For this reason, guaranteeing
observational sessions up to few hours before re-entry is always
recommended to reduce the size of re-entry windows. We also attempted
to generalize the results obtained for GOCE to similar simulated
orbits for objects of different shapes (spherical or cylindrical).

\begin{figure}[h]
\begin{center}
\includegraphics*[width=3cm]{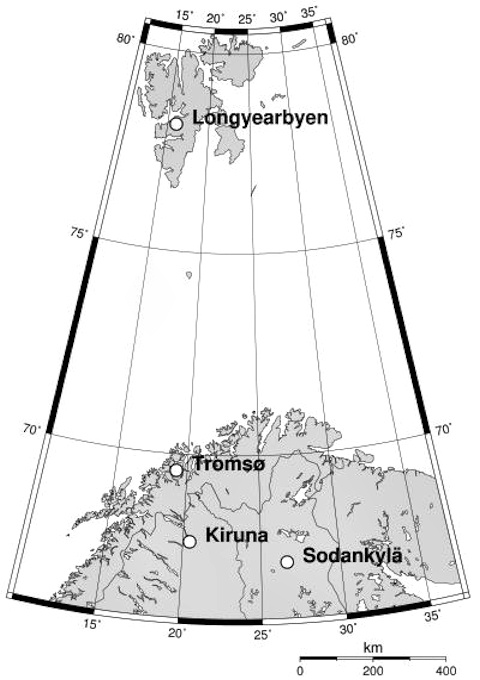}
\hspace{2cm}
\includegraphics*[width=6cm]{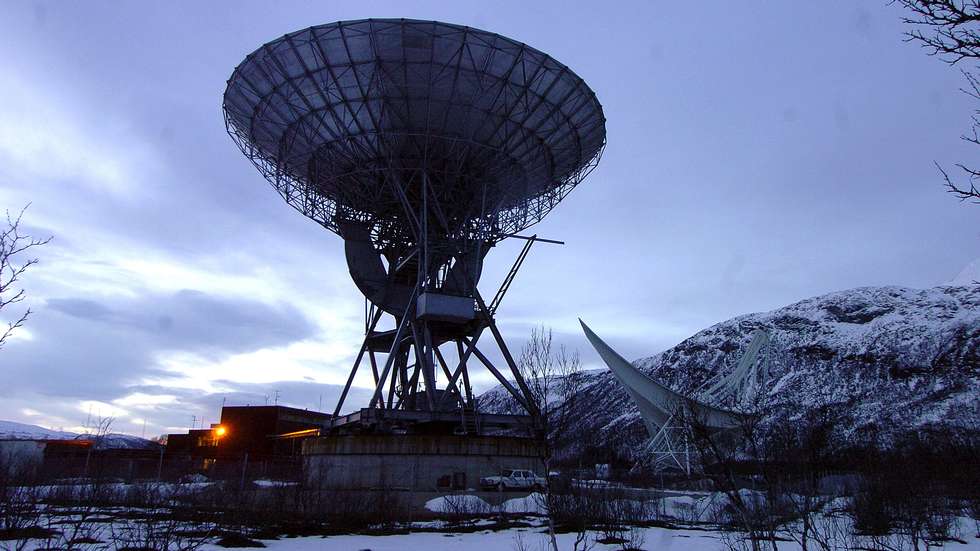}
\end{center}
\caption{Left: EISCAT facilities map. Right: EISCAT UHF radar at Troms\o}\label{fig:eismap}
\end{figure}

The aim of the present work is to test, via numerical simulations, and
also with real data processing, the application of a particular
radar-based tracking approach, described in Section~\ref{sec:eis_uhf},
for re-entry predictions. This technique consists in the generation of
very short tracking passes, typically of few seconds, of a decaying
object, to be exploited for OD and ballistic coefficient
estimation. In other words, the main purpose of this task is to assess
to which extent very short tracks from a high-latitude (polar) radar
station can be useful to improve the re-entry predictions of a
decaying object, at least for a GOCE-like circular polar orbiter. It
is well known that, in many cases, the main source of information
during re-entry campaigns are the US Two Line Elements (TLE), for
which no related uncertainty information is provided. Thus, it is
desirable for an institution such as the European Space Agency to have
its independent observational resources. An important goal to be
achieved in the near future is to free ESA from the dependency from US
TLEs release for re-entry campaigns, as much as possible. A clear step
in this direction is not only the increased availability of the german
TIRA radar, but also the exploitation of european instruments
originally conceived for different purposes such as the EISCAT
(European Incoherent Scatter Scientific Association, see
Figure~\ref{fig:eismap}) facilities, for the observation of decaying
objects, and space debris in general (see e.g. \cite{netal12}). It is
straightforward that these European stations are to be preferred with
respect to other, less affordable, instruments. The possibility to
extensively use the very short tracks of the EISCAT sensors in
place of other instruments for re-entry predictions shall be examined
in this work, by a comparison of the results obtained from comparable
scenarios which include and do not include the short tracks.

This problem can be analyzed from at least three points of view:
\emph{(i)} frequency of re-entry predictions, \emph{(ii)} drag
coefficient estimation, \emph{(iii)} optimization of resources. These
three items are strongly related to each other. A better drag
coefficient estimation can be achieved with a higher frequency of
observations, which on the one hand corresponds to a higher frequency
of possible re-entry predictions, and on the other hand implies a
larger exploitation of resources. In general, while during the first
part of the decay phase few observational resources should be enough
to obtain acceptable re-entry predictions, during the very last part
of re-entry (e.g. last 48 hours) it is highly recommended to provide
more frequent observational data.

Originally designed for observing the ionosphere by means of the
incoherent scatter method, the EISCAT radar sites are more suitable to
observe polar orbiters rather than low inclination ones, due to their
high latitude location. The main features of the four EISCAT sites are
given in the following list:
\begin{enumerate}
\item Troms\o (Norway): UHF Transmitter/Receiver at $\sim 928$MHz (a $32$m
  steerable parabolic dish), and VHF Transmitter/Receiver at $\sim 224$MHz
  (four $30\times 40$m steerable parabolic cylinder);
\item Kiruna (Sweden): VHF Receiver at $\sim 227$MHz (a $32$m steerable
  parabolic dish);
\item Sodankylä (Finland): VHF Receiver at $\sim 227$MHz (a $32$m steerable
  parabolic dish);
\item Longyearbyen (Norway): UHF Transmitter/Receiver at $\sim 500$MHz (a
  $32$m steerable parabolic dish, and a $42$m fixed parabolic dish).
\end{enumerate}
In the following, we will focus on the EISCAT UHF radar working at
$930$MHz frequency located in Troms\o, Norway.

\section{EISCAT UHF sensor and main features}
\label{sec:eis_uhf}

Three important references concerning the employment of the EISCAT
sensors for space objects observations are \cite{netal12},
\cite{metal13} and \cite{vk17}. In particular, the sensor which has to
be dedicated mainly to tracking observations is the EISCAT UHF radar
working at $930$MHz frequency located in Troms\o, Norway. Due
to slow antenna motors not capable to smoothly track a target, the
radar measurement strategy is based on a scheduled ``point-and-steer''
technique, and the peculiarity of this tracking system resides in the
very short length of the measured observation sets, of few
seconds. This is quite different from standard radar systems
techniques (e.g. TIRA) that typically produce continuous observational
tracks of several minutes \cite{letal14}.

In order to be able to test the capabilities of such a particular
measurement system for re-entry predictions, we shall need to identify
the general characteristics of the instrument and of the technique,
and to appropriately traduce them under the form of \emph{computed
observables} to be implemented in our Simulation and Orbit
Determination software.

\begin{figure}[h]
\begin{center}
\includegraphics*[width=6.5cm]{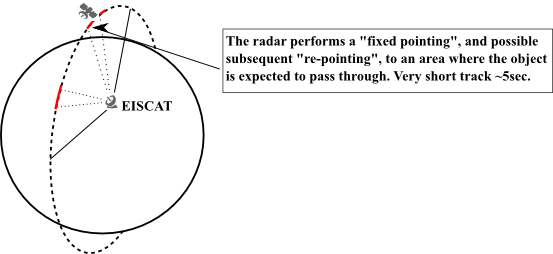}
\includegraphics*[width=6.5cm]{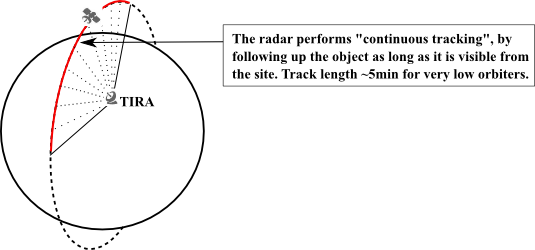}
\end{center}
\caption{Left: EISCAT limited tracking capabilities. Right: TIRA
  standard tracking capabilities.}\label{fig:eistracking}
\end{figure}

According to \cite{vk17}, we can identify and define the
following main features:
\begin{enumerate}
\item Location of the sensor $69.58649229^\circ N$, $19.22592538^\circ E$, $85.554$m
\item Measurement types and expected accuracies Range - approximately
  $15$m $1\sigma$ errors (as two-way measurement); Range-rate (Doppler
  shift) - approximately $1$m/s $1\sigma$ errors (as instantaneous
  measurements); Azimuth/Elevation - no accurate measurement is
  possible (no mono-pulse feed available).
\item Visibility conditions and measurement frequency Elevation
  threshold $30^\circ$; Dish steering speed $\sim 1.3^\circ$/s, the
  antenna controller cannot smoothly track targets, it can only move
  to a position and stop there.
\end{enumerate}
The system can be used for limited tracking observations of targets
with approximately known orbital elements. The tracking is done using
a technique which involves poiting to a location where the target will
approximately be and wait for it to pass. Once a preliminary orbital
information is available, for example from TLE or other sources, it is
possible to estimate the maximum number of points that the antenna can
keep up with, i.e. that it is able to steer, point and wait until the
object passes through the beam, within the total time span during
which the object is potentially visible over the sensor (elevation
greater than $30^\circ$). In \cite[Fig.2.6 and Fig.2.7]{vk17} there
are two examples, showing that the antenna is capable to keep up with
one steer-and-point every $\sim 1$ minute, and that the object is
expected to pass through the beam for very few seconds. As a result,
we have a single measurement ``spot'' per fixed pointing position
consisting of few seconds range and range-rate
data. Figure~\ref{fig:eistracking} is a graphical representation of
this technique, in comparison with standard countinous smooth tracking
of few minutes duration.

For objects with an orbit high enough to stay over the station for
several minutes (e.g. $5$min) it is possible to perform multiple
observation spots and to form a tracklet for orbit determination
purposes, which can be comparable to a continous tracking
pass. However, this is not possible for very low orbiters, such as
GOCE or any low circular polar orbiter during decay phase. We will see
in the following that for these objects the visibility contraints are
such that it is possible to get only one spot per pass.

In this work, we need to focus on re-entry predictions, and in
particular on low circular polar orbiters and their ballistic
coefficient calibration. A basic comparison with standard radar
systems' features at this point is mandatory. According to
\cite{letal14}, \cite{bvetal14}, and as reported in \cite{cetal17},
conventional radar systems can be effectively used for re-entry
predictions calibrations. Depending on the geographical location of
the radar site, and on the orbital features of the tracked object,
different visibility conditions occur. What we can assume for
acceptable visibility conditions is a minimum elevation of the
detected object over the local horizon of at least $2-3^\circ$, on condition
that enough information for the calibration of atmospheric refraction
is available. For instance, the typical length of a tracking pass is
of few minutes for very low orbiters, while the frequency of
visibility of subsequent passes from the same station strongly depends
on its latitude. Tracking passes which reach a higher elevation peak
are preferable, e.g. greater than $10^\circ$, since they usually contain more
information on the orbit, the atmospheric refraction error is lower,
and also provide a lower minimum range distance. As regards the
nominal accuracies that can be achieved by standard tracking
instruments, values for measurements error's RMS (noise) around $10-30$m
for range and $0.01-0.03^\circ$ for azimuth/elevation are commonly considered
acceptable to perform good re-entry predictions calibration. In fact,
even better values for both measurement's noise and bias are in
principle achievable with high quality radar like TIRA, but extended
body uncertainties must be taken into account for large objects, and
the presence of large dynamical model's errors at very low altitudes
ends up to dominate the measurement’s errors over long time spans.

In conclusion, the main difference between EISCAT UHF sensor and the
standard radar considered so far, resides in the high elevation
threshold, in the lack of continuous tracking capabilities and in the
low accuracy direction angle information. We will analyse in the
following sections how these features can affect the orbit
determination and ballistic coefficient calibration problems, and to
which extent we can effectively obtain useful information for re-entry
predictions.


\section{Examples from 2012-006K AVUM R/B real data}
\label{sec:2012-006K}

As reported in \cite{vk17}, several experimental tracking observations have
been performed with EISCAT, on selected targets, to test EISCAT UHF
capabilities for orbital elements determination. As mentioned before,
the tracking program utilizes already available TLE orbital elements
to plan an antenna scan for tracking of objects. In particular, the
selected target 2012-006K VEGA AVUM R/B is of special interest for our
study, because it was a re-entry object.

\begin{figure}[h]
\begin{center}
\includegraphics*[width=5cm]{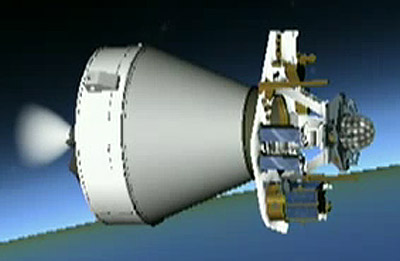}
\end{center}
\caption{2012-006K AVUM R/B.}\label{fig:avum}
\end{figure}

According to the ESA DISCOS database, the 2012-006K AVUM rocket body
had a nominal mass of $960$kg, with an approximate shape of $1.9\times 1.7
\times 1.9$m$^3$, and an average cross sectional area of $2.162$m$^2$ (Figure~\ref{fig:avum}).

The object re-entered on 2016-11-02. We show on Table~\ref{tab:avum} the TLE
osculating orbital elements on 2016-10-13, 20 days before re-entry,
when the minimum altitude of the object was around $210$km.
\begin{table}
\caption{TLE 2012-006K osculating keplerian elements on 2016-10-13 at 19:48:43 UTC, in GCRS}
\begin{tabular}{llllll}
\hline
\emph{a} & \emph{e} & \emph{i} & \emph{$\Omega$} & \emph{$\omega$} & \emph{M} \\
\hline
$6649.594\,$km & $8.1817 \times 10^{-3}$ & $69.3784^\circ$ & $246.4754^\circ$ & $176.8664^\circ$ & $183.2227^\circ$ \\
\hline
\end{tabular}
\label{tab:avum}
\end{table}

Within this work, ESA provided us with one TIRA pass and four EISCAT
passes for 2012-006K close to re-entry. Main information on passes are
gathered in Table~\ref{tab:avum_passes}.
\begin{table}[h]
\caption{Summary of TIRA (T1) and EISCAT UHF (E1-E4) tracking passes provided by ESA as CFI}
\begin{tabular}{llll}
\hline
Pass & Obs. type & Data start UTC & Data end UTC \\
\hline
T1 & $\rho$,$\dot{\rho}$,$Az$,$El$ & 2016-10-20 14:50:06&2016-10-20 14:56:09 \\
E1 &$\rho$,$\dot{\rho}$ & 2016-10-21 16:19:48.98&2016-10-21 16:19:50.38 \\
E2 &$\rho$,$\dot{\rho}$ &2016-10-22 16:09:53.32 &2016-10-22 16:09:55.50 \\
E3 &$\rho$,$\dot{\rho}$ & 2016-10-22 19:12:27.81&2016-10-22 19:12:29.88 \\
E4 &$\rho$,$\dot{\rho}$ &2016-11-01 15:59:48.26 &2016-11-01 15:59:50.00 \\
\hline
\end{tabular}
\label{tab:avum_passes}
\end{table}

By combining the TIRA and EISCAT passes in different ways we can
perform OD and ballistic coefficient calibration, in order to obtain
different re-entry predictions. The main dynamical assumptions for the
data processing consists in
\begin{itemize}
\item a static gravity field up to degree and order 16, 
\item luni-solar gravitational perturbations, 
\item a non gravitational drag acceleration:

\begin{equation}
 \mathbf{a}_{drag} = -\frac{1}{2} B \rho v_r^2 \hat{\mathbf{v}}_r  
\label{eq:drag_acc}
\end{equation}
where $\rho$ is the atmospheric density at the object's location,
modeled with NRLMSISE00, using daily and 81d averaged F10.7 solar flux
indices, and daily and 3h Ap geomagnetic indices as space weather
proxies. $B=C_dA/m$ is the ballistic coefficient, assumed to be
constant over a single propagation (3DOF approach), $A$ is the
reference area of the spacecraft, $C_d$ is the drag coefficient, $m$
is the satellite mass, $\hat{\mathbf{v}}_r$ is the direction of the
relative velocity of the spacecraft with respect to the atmosphere and
$v_r=|\mathbf{v}_r|$. For the relative velocity computation, we assume
that the atmosphere is rotating fixed with the Earth
\cite{mg05}.
\end{itemize}
In this case, the preliminary information for the orbital state
solution (first guess) is always based on the available TLEs,
tipically chosen at an apoch close to the middle of the observational
interval, while a preliminary value for the drag coefficient $C_d$ is
around 4.1, for a total of 6+1 parameters to be determined in each OD
solution (6 initial conditions and 1 ballistic coefficient).

We have combined three re-entry prediction scenarios, summarized in Table~\ref{tab:avum_scen}.
\begin{table}[h]
\caption{Re-entry prediction scenarios for 2012-006K with the available radar tracks.}
\begin{tabular}{llll}
\hline
scen & passes & total obs. $\Delta T$ & residual lifetime from last obs \\
     &        &                       &     to nominal re-entry (nov-2) \\
\hline
1 & T1 + E1  & $\sim 25$h & $\sim 12$h\\
2 & E1 + E2 + E3 & $\sim 27$h  & $\sim 11$h \\
3 & T1 + E1 + E2 + E3 & $\sim 52$h & $\sim 11$h \\
\hline
\end{tabular}
\label{tab:avum_scen}
\end{table}

The main results for the three re-entry prediction scenario are described in Table~\ref{tab:avum_res}.
\begin{table}[h]
  \caption{Summary of main results for TIRA + EISCAT UHF re-entry predictions of 2012-006K. AP stands for the application of an A-Priori constraint on the initial conditions of 1km in position and 1m/s in velocity (TLE-based). NM stands for Not Mandatory, i.e. the OD scenario is stable.}
\begin{tabular}{lllll}
\hline
scen & AP & residuals RMS & estim. $C_dA$ & re-entry epoch \\
     &        &                       &    & UTC (at 90km) \\
\hline
1 & Y     &T1($\rho$,$Az$,$El$): $10.4$m, $0.0059^\circ$, $0.0067^\circ$  & $8.8373\,$m$^2$  & 11-03 $\sim$19:56 \\
  & (NM)  & E1($\rho$,$\dot{\rho}$): $11.7$m, $1.8$m/s                                     &  & \\
\hline
2 & Y  & E1($\rho$,$\dot{\rho}$): $11.1$m, $1.8$m/s & $8.7332\,$m$^2$  & 11-03 $\sim$22:00 \\
  &   & E2($\rho$,$\dot{\rho}$): $11.5$m, $1.1$m/s &                  &                  \\
  &   & E3($\rho$,$\dot{\rho}$): $9.6$m, $1.0$m/s&                  &                   \\
\hline
3 & Y  &T1($\rho$,$Az$,$El$): $13.5$m, $0.015^\circ$, $0.014^\circ$   & $9.0684\,$m$^2$  & 11-03 $\sim$11:51 \\
  & (NM)  & E1($\rho$,$\dot{\rho}$): $66.8$m, $4.5$m/s &                  &                  \\
  &   & E2($\rho$,$\dot{\rho}$): $28.8$m, $11.5$m/s &                  &                   \\
  &   & E3($\rho$,$\dot{\rho}$): $17.7$m, $6.3$m/s &                  &                  \\
\hline
\end{tabular}
\label{tab:avum_res}
\end{table}

As we can see from Table~\ref{tab:avum_res}, the scenario 2, that does
not include the TIRA pass, needs an a priori assumption on the initial
conditions error to be included in the fit to obtain stable
convergence. In this case the assumption consists in a 1km 3D position
and 1m/s 3D velocity a priori covariance matrix, which may be
reasonable for the TLE first guess state. This can be understood if we
consider that one TIRA pass consists of $\sim 5$ minutes of accurate range,
azimuth and elevation measurements, thus containing a quite
significant amount of orbital information, while one EISCAT pass
basically consists of only one ``spot'' of range and range-rate
data. Since the full OD solution has 7 solve-for parameters, this
means that only three EISCAT tracks are not enough to have a full
solution with stable differential corrections, and even though in both
scenarios 2 and 3 we used the same TLE as first guess for the initial
conditions, the case 2 contains too few data and additional
information on the orbit is needed. This will be discussed in more
details in the following Section~\ref{subsec:crit_cases}. 

As regards the fourth EISCAT track on 2016-11-01, on the one hand it
is very close to re-entry and on the other hand it is very far from
the previous observations available ($\sim 10$ days). We believe that
a calibration with only these data and over a such long time span
would be a quite unrealistic and unlucky situation. For a re-entry
prediction to be computed during the very last day of decay, a
calibration time span at the order of less than one day should be more
suitable. In order to try to fit this radar track as it would have
been done in reality, we can define a re-entry prediction scenario
mixing TLEs and the EISCAT track (Table~\ref{tab:avum_scen4}). A
standard weight of 1km and 1m/s is applied to the fitted TLEs (see
also Section~\ref{sec:tle_calib}).
\begin{table}[h]
  \caption{Re-entry prediction scenarios for 2012-006K with the last available radar track and two TLEs.}
\begin{tabular}{llll}
\hline
scen & passes & total obs. $\Delta T$ & residual lifetime from last obs \\
     &        &                       &     to nominal re-entry (nov-2) \\
\hline
4 & TLE1(10-31$\sim$20:32) + & $\sim 19$h  &  $\sim 12$h  \\
  & TLE2(11-1$\sim$03:53) + E4  &             &           \\
\hline
\end{tabular}
\label{tab:avum_scen4}
\end{table}

The main results for the three re-entry prediction scenario are
described in Table~\ref{tab:avum_scen4_res}.
\begin{table}[h]
\caption{Summary of main results for TLE + EISCAT UHF re-entry predictions of 2012-006K}
\begin{tabular}{llll}
\hline
scen & residuals RMS & estim. $C_dA$ & re-entry epoch \\
     &                         &    & UTC (at 90km) \\
\hline
4 & TLEs(pos,vel): $170$m, $0.2$m/s  & $9.5235\,$m$^2$  & 11-02$\sim$04:57   \\
  &E4($\rho$,$\dot{\rho}$): $12.9$m, $1.8$m/s      &    & \\
\hline
\end{tabular}
\label{tab:avum_scen4_res}
\end{table}

\section{Re-Entry predictions simulations with GOCE POD}
\label{sec:gocesim}

In order to perform a deeper analysis on the performances and
capabilities of the particular radar measurement system previously
introduced, we can set up a reliable simulation scenario, as we have
done in \cite{cetal17} for the GOCE spacecraft and other simulated
objects. The test case scenario that we want to consider shall
consist of a dynamical scenario, i.e. the orbiting object's main
features, and an observational scenario, i.e. the set of data
available. We are going to focus on the typical case of a high
inclination (polar) LEO in an almost circular orbit at very low
altitude, which is decaying to re-entry in few weeks or days.

The first thing to be defined is the list of dynamical assumptions for
the orbit of the object that we are going to consider and the forces
acting on it, e.g. consisting in a static gravity field up to a
suitable degree and order, planetary perturbations, drag perturbations
with a ballistic coefficient to be estimated during the orbit
determination process.  The second step in the definition of the tests
is the definition of the observational scenarios. In order to be able
to perform a sensitivity analysis with respect to different
situations, which means to different availability of observational
data, we will need to consider several possibilities in terms of
active/inactive radar sites, released TLEs, and additional very short
tracks (VST). Thus, a scenario would consist of a set of radar
tracking passes, TLEs and whatever additional useful measurement on
the orbiting object. To each particular scenario corresponds an orbit
determination and a ballistic coefficient estimation, necessary to
calibrate a future re-entry prediction. A ``re-entry campaign'' can
thus be simulated by considering a set of subsequent observational
scenarios with observations available up to the very last orbits
before the object's re-entry in the lower atmosphere.  As extensively
done in \cite{cetal17} for the case of single- vs. multi-sensor
scenarios, here we want to perform the same kind of tests to compare
the cases of observational scenarios including and not including
EISCAT's very short tracks for re-entry predictions. In practice, two
different scenarios are considered comparable if they have the same
epoch of prediction, i.e. if their last observational data available
is more or less at the same epoch. Given two comparable scenarios, we
are able to perform a standard OD and ballistic coefficient estimation
with the data available for each of them. Two resulting re-entry
predictions can be computed, which may be significantly different
mainly depending on the total time span of the chosen observations and
on the atmospheric model adopted. To the purpose of a reliable and
correct comparison between different scenarios, we can make use of the
following quantities: the nominal predicted re-entry epoch
(conventionally at 90km altitude), the absolute and relative
re-entry time error (by exploiting a priori information) with respect
to the total re-entry residual lifetime, and the post-fit measurements
residuals. Moreover, if enough information is available on the orbit
of the re-entering object, then such an orbit can be used as a
reference for comparison with the fitted OD solution, and as a source
to extrapolate a refined ballistic coefficient's behavior, e.g. with a
suitable piece-wise constant (PWC) function. For example reference orbits
can be the accurate GPS-based POD in the GOCE case \cite{getal14}, the
simulated orbit if the trajectory is generated within the whole test
process, or just the released TLE states (see Section~\ref{sec:tle_calib}). The behavior
of the ballistic coefficient obtained from such richer amount of
information on the decaying object, can then be compared with the one
obtained from the radar observational scenarios defined before, which
contain only sparse data from different sources.

We have already shown in \cite{cetal17} that, provided a reasonable
frequency of observations, radar-based re-entry predictions
calibrations are very effective in capturing the average drag effect
over the time span covered by the measurements. The most important
lesson learned from that study was the evidence of the dominance of
dynamical model's errors at decaying altitudes (typically 100-300km),
due for instance to the difficulties in having very accurate
atmospheric density models available and in accurately predicting
significant attitude variation. Errors at the order of 10-20\% are
commonly found, a particular density model can be better than another
for a particular range of altitudes, or for a particular period of
time, depending on solar flux and geomagnetic storms events, but none
of them can be considered always better than the others
\cite{petal04}. The differences between two given density models over
a fixed time span are not only given by a simple constant bias. For
this reason, even if the ballistic coefficient estimation does take
into account the correction of a constant bias over a fixed time span,
and this correction is quite effective in absorbing part of the
mis-modeled dynamics, this bias is not enough to compensate
for all the errors in the atmospheric density, both on shorter time
intervals and over the time span from the epoch of prediction to the
actual re-entry. Thus, the adoption of different density models could
in principle lead to significantly different re-entry
predictions. Finally, particular changes in the attitude motion of
elongated, or just asimmetric, objects could lead to large variations
of the average ballistic coefficient and thus of the re-entry time.

\subsection{Main assumptions}
\label{subsec:assumptions}

Also in this case, the main dynamical assumptions for the data
processing consists in: a static gravity field up to degree and order
16, luni-solar gravitational perturbations, a 3DOF non gravitational
drag acceleration with atmospheric density model NRLMSISE00, using
daily and 81d averaged F10.7 solar flux indices, and daily and 3h Ap
geomagnetic indices as space weather proxies.  Reference
\cite{cetal17} holds for the assumptions in the simulation of the TIRA
radar measurements: elevation threshold of $2^\circ$, random generated
noise of 10m in range and $0.01^\circ$ in azimuth and elevation, using
the GOCE POD as reference trajectory. The simulated TIRA passes which
have a maximum elevation peak lower than $10^\circ$ are generally not
used here. No addition of tropospheric refraction effects is
considered in the simulations, assuming that they are already removed
to a level lower than the considered noise (see
e.g. \cite{mg05}). A deeper dedicated analysis on this
latter topic is considered beyond the purposes of this work.

As regards the specific EISCAT UHF measurements simulation, we need to
take the cue from what is described in \cite{vk17}, and summarized in
Section~\ref{sec:eis_uhf}. Given an elevation threshold of $30^\circ$,
we assume to be able to schedule an antenna scan per minute of
visibility over the station. In the case of a low orbiter such as
GOCE, the duration of a visibility from EISCAT over $30^\circ$ of elevation
is of the order of one minute (see results below). This implies that
in general we can assume to be able to point the radar around the
middle of the expected pass, and wait for the object to pass through
the beam. The duration of a satellite passing through the radar beam
is a few second, we assume to take 2 seconds of range and range-rate
measurements, with a frequency of 0.1s. Note that a shorter frequency
of 0.01s could also be used, but such track is so short that we expect
it to be more or less equivalent to a single measurement of range and
range-rate. The simulated noise for such measurements is of 15m in
range and 1m/s in range-rate. Finally, no observation of azimuth and
elevation is used, even if we believe that at least a low accuracy
information just from antenna pointing could be useful in some cases
(see Section~\ref{subsec:crit_cases}).

\begin{figure}[h]
\begin{center}
\includegraphics*[width=14cm]{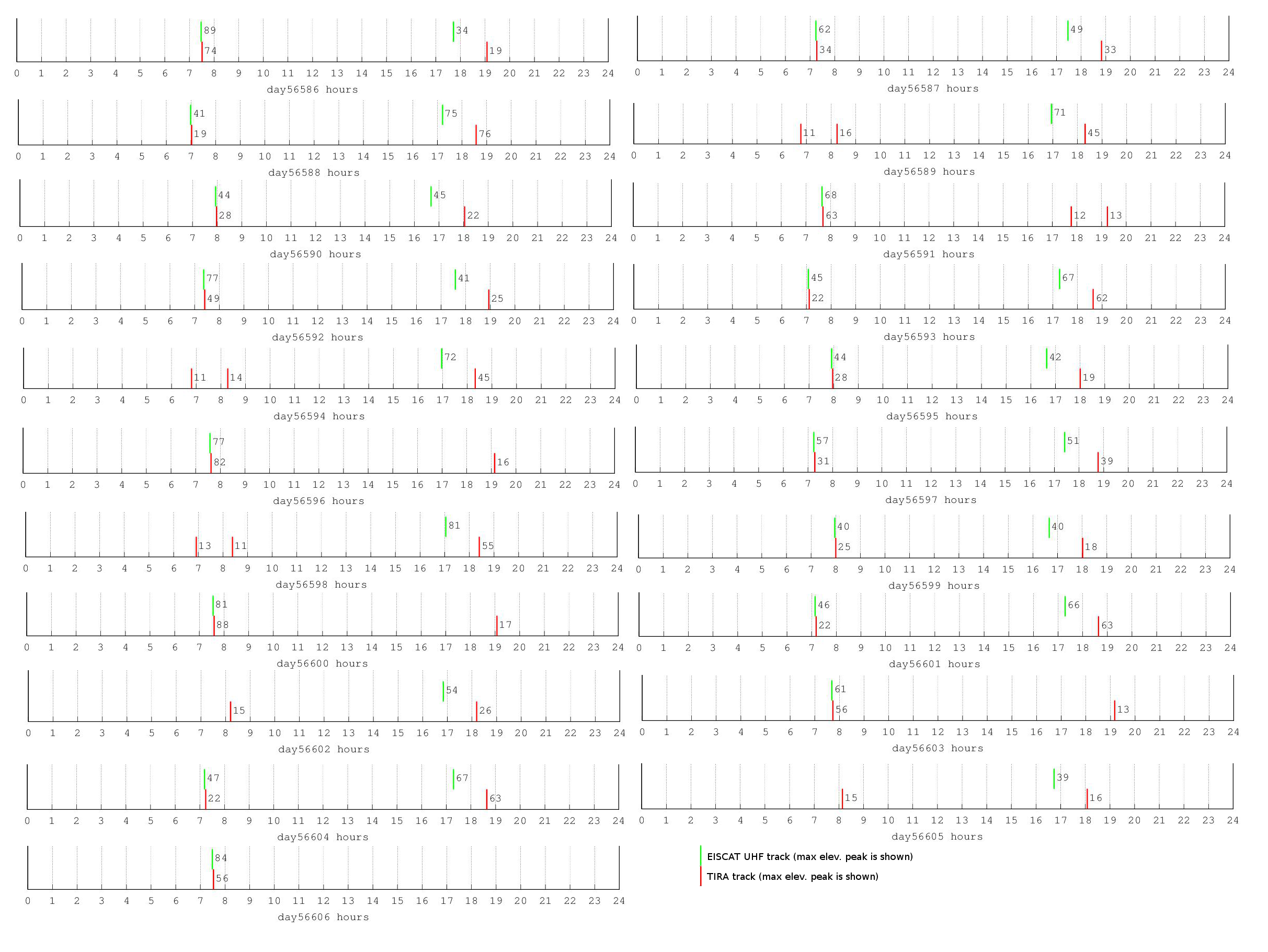}
\end{center}
\vspace{-1cm}
\caption{Visual representation of tracking passes from TIRA (red) and
  EISCAT UHF (green) during the three weeks of GOCE decay (time is in
  MJD). Maximum elevation peaks of passes are shown in deg, discarding
  passes with elevation peaks of less than 10$^\circ$.}\label{fig:passes}
\end{figure}

A visual representation of the visibility of GOCE from TIRA and EISCAT
UHF is given in Figure~\ref{fig:passes}, for the three weeks of decay
(from MJD 56586 to MJD 56606). The maximum elevation peaks are also
shown. The average duration of the TIRA passes is $\sim 5.8$min, while the
average duration of the visibility over the EISCAT UHF radar is
$\sim 1.3$min. As already stated, we assume to be able to get just one spot
of $\sim 2$ seconds per pass, around the middle of each pass. Another
drawback of the high elevation threshold of $30^\circ$ is the possibility to
have larger gaps of 24 hours between two subsequent observations. It
is interesting to note that the EISCAT and TIRA tracks are very close
together. The EISCAT sensor has a higher latitude of $\sim 20^\circ$ with respect
to TIRA, and a different longitude of $\sim 12^\circ$. In principle, due to the
higher latitude, without the constraints on the elevation threshold
and on the minimum elevation peak, the EISCAT sensor would allow for
much more visibility and a much higher frequency of passes. However,
in every case the geometrical conditions turned out to be such that
the EISCAT and TIRA passes with the highest elevation peaks are always
not more distant than one orbit, and these are also the only ones that
satisfy the former elevation constraints (shown in Figure~\ref{fig:passes}).

The basic idea is the same as the one described in \cite{cetal17} for the
case of TIRA-only and many sensors re-entry prediction scenarios. The
reference trajectory (GOCE POD) is fully exploited to estimate a
refined piecewise constant ballistic coefficient that is able to
compensate the mis-modeled non gravitational perturbations up to a
certain level of accuracy. A suitable subinterval of 30min was
considered a good value to reconstruct the variations the ballistic
coefficient from the orbit, but also shorter subintervals were
possible. We showed that with a station like TIRA, OD and
ballistic coefficient calibration with radar passes over 24-48h are
very effective in estimating the average behavior of the ballistic
coefficient and a good re-entry trajectory, in comparison with the
30min PWC coefficient and with the POD orbit itself. As we approach
the last part of re-entry, calibrations over shorter time intervals
(e.g. 12h) are also effective.

\begin{figure}[h]
\begin{center}
\includegraphics*[width=8cm]{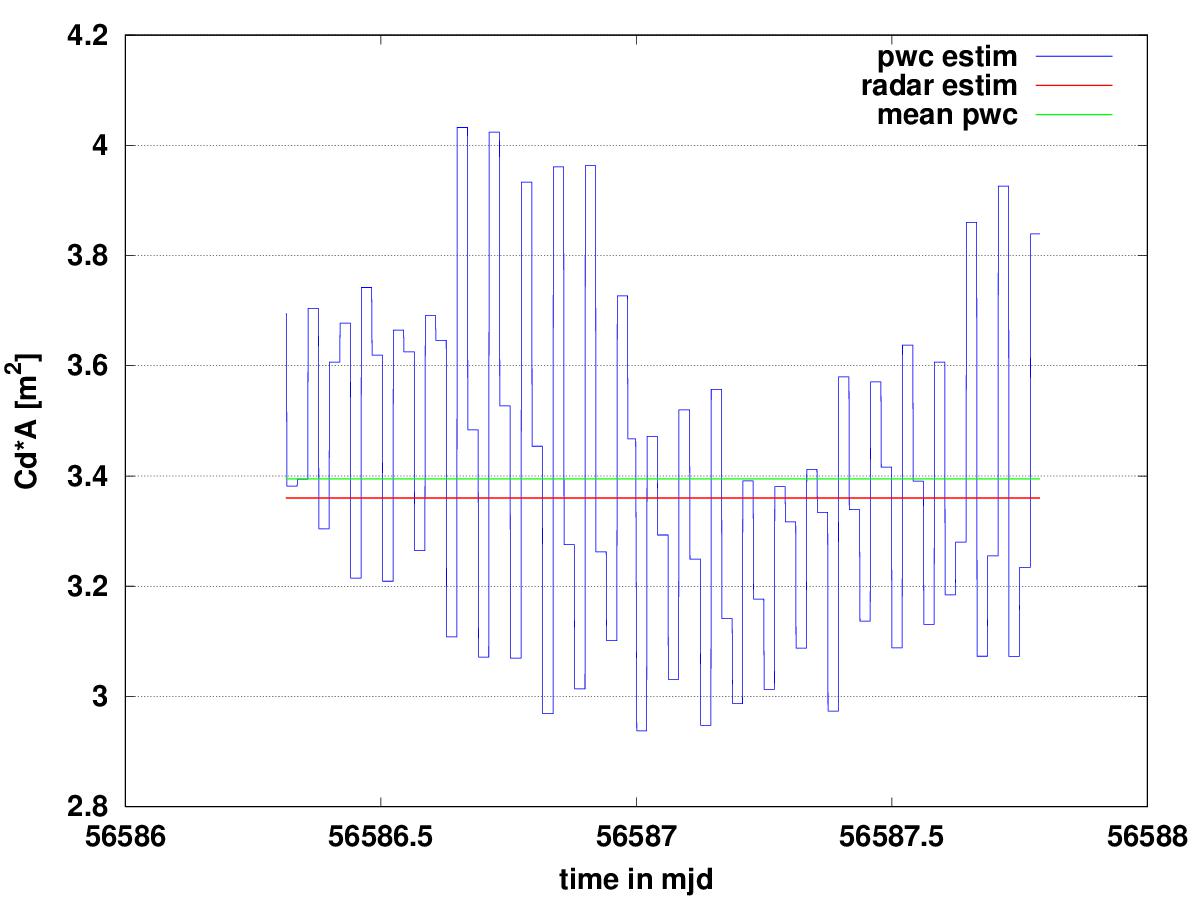}
\end{center}
\caption{TIRA-based ballistic coefficient estimation (in red) over
  re-entry days 1 and 2, compared with POD-based 30min PWC coefficient
  (in blue). Relative difference with respect the mean value (in
  green) is $\sim 1\%$.}\label{fig:tira_cda_estim_sc1}
\end{figure}

As an example, we show in Table~\ref{tab:goce_scen1} the results for the OD scenario
which includes the first four TIRA passes during re-entry days 1 and
2, in terms of comparison with the reference orbit. In Figure~\ref{fig:tira_cda_estim_sc1} we can
see the estimated ballistic coefficient in comparison with the
POD-based 30min PWC coefficient. The corresponding estimated re-entry
epoch is Nov-11 at $\sim$11:24 UTC.
\begin{table}[h]
  \caption{RMS differences between TIRA-based estimated orbit and POD over re-entry days 1 and 2, in the R-T-W system and in osculating keplerian elements semimajor axis $a$, eccentricity $e$, and mean argument of latitude $\omega+M$}
\begin{tabular}{llllllllll}
\hline
    &$\Delta \mathbf{x}_R$ &$\Delta \mathbf{x}_T$ &$\Delta \mathbf{x}_W$ & $\Delta \mathbf{v}_R$ &$\Delta \mathbf{v}_T$ &$\Delta \mathbf{v}_W$ \\
\hline
RMS & 192.0m &542.7m &81.7m &0.5m/s &0.2m/s &0.1m/s \\
\hline
  & $\Delta a$ & $\Delta e$ & $\Delta (\omega+M)$ \\
\hline
RMS & 10.3m &$3.4\times 10^{-5}$ & $0.0032^\circ$ \\
\hline
\end{tabular}
\label{tab:goce_scen1}
\end{table}

\begin{figure}[h]
\begin{center}
\includegraphics*[width=8cm]{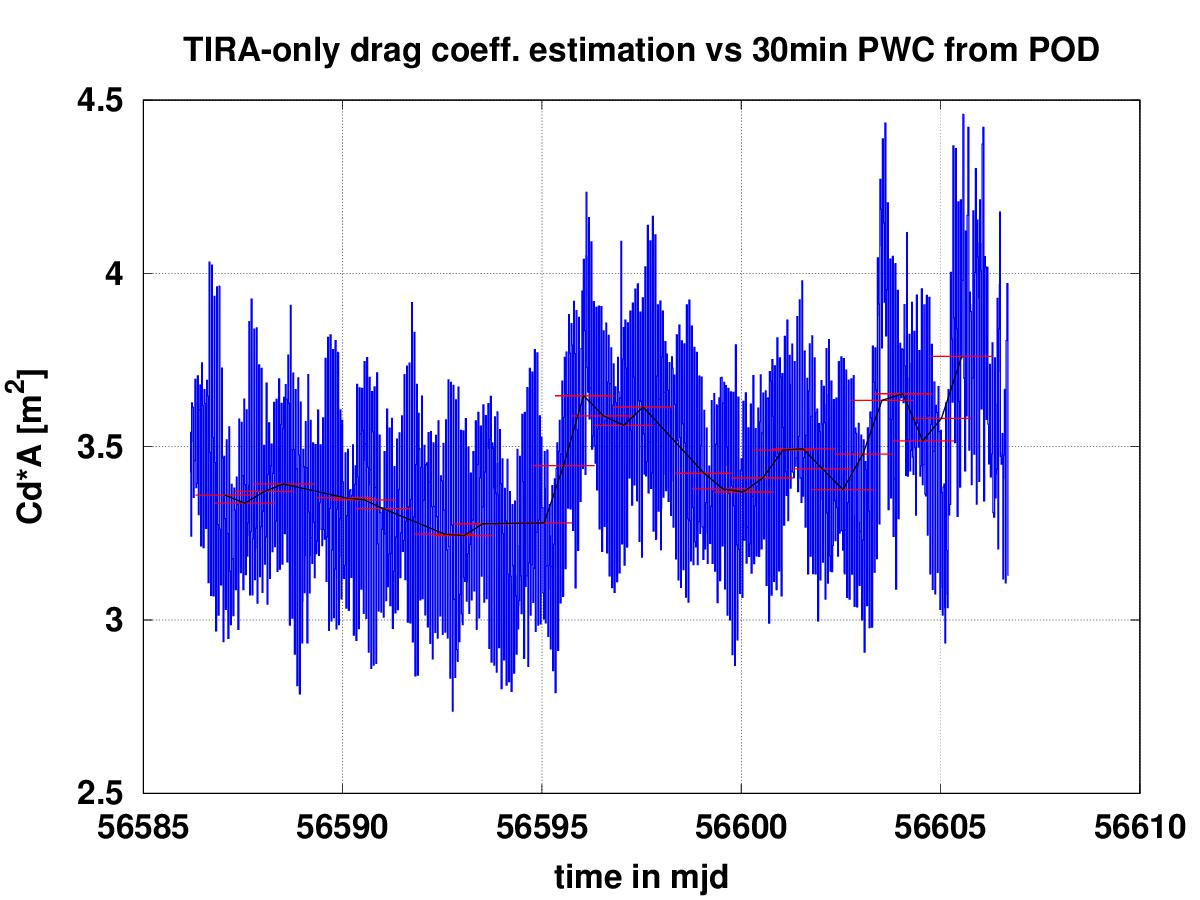}
\end{center}
\caption{TIRA-only ballistic coefficient estimation over intervals of
  ~36 hours (in red), compared with 30min PWC coefficient from POD (in
  blue).}\label{fig:tira_only_drag}
\end{figure}

\begin{figure}[h]
\begin{center}
\includegraphics*[width=6.5cm]{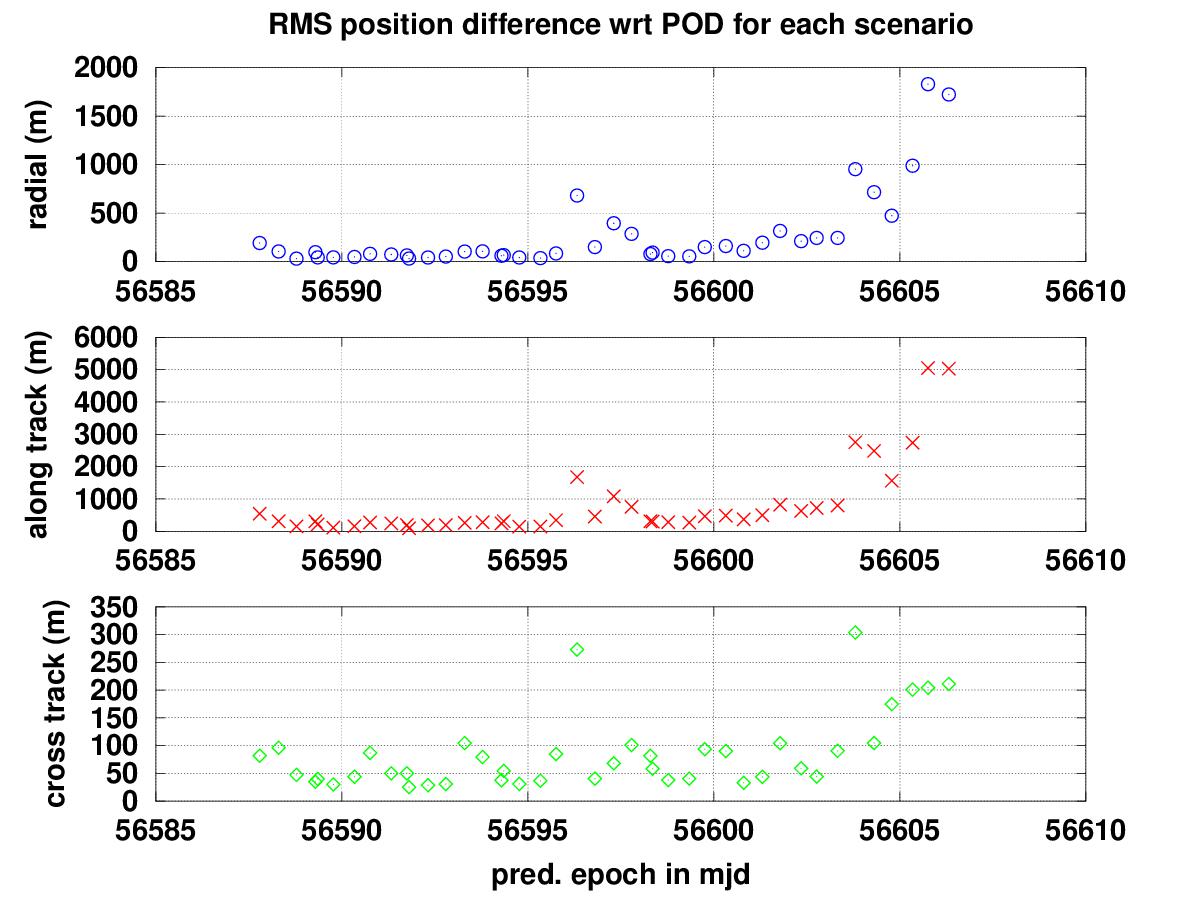}
\includegraphics*[width=6.5cm]{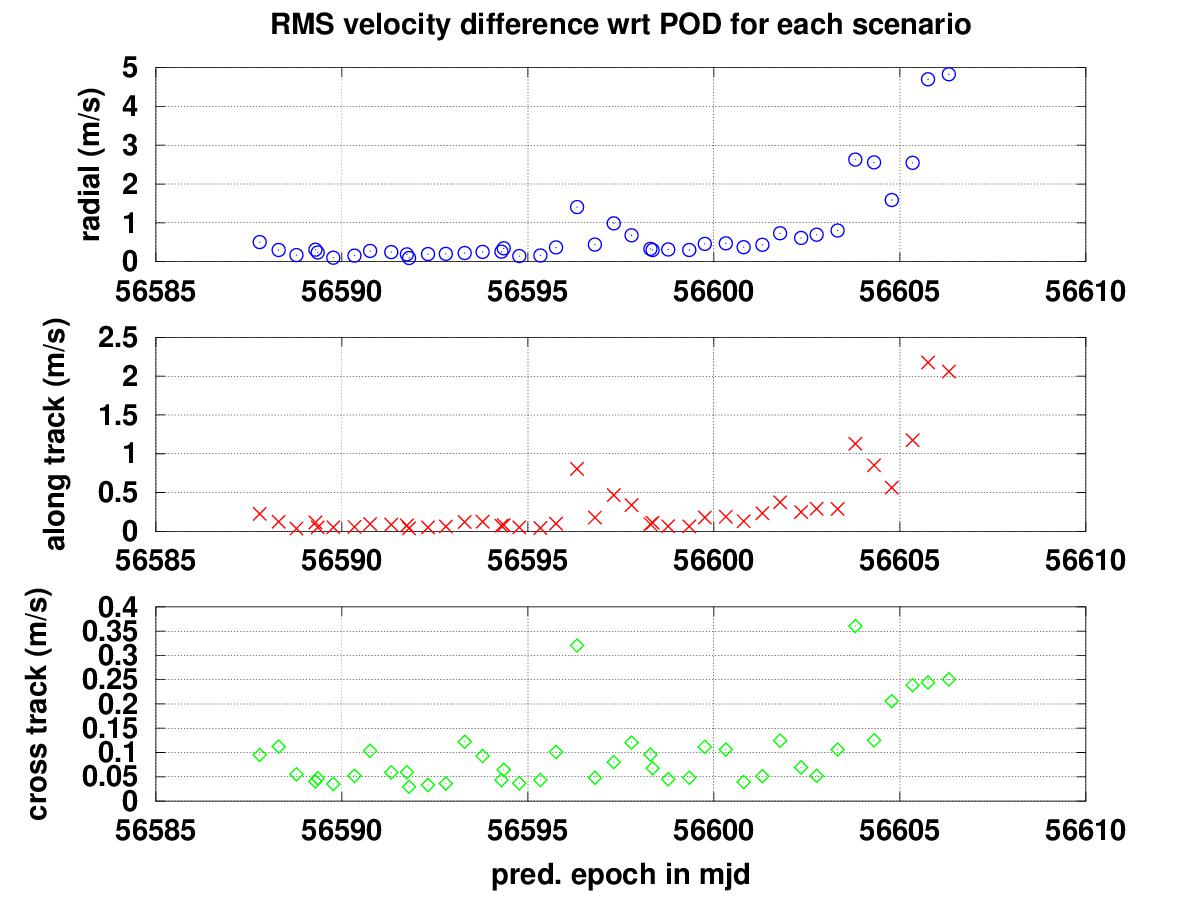}
\end{center}
\caption{RMS of position (left) and velocity (right) differences
  w.r.t. POD for each TIRA-based prediction scenario.}\label{fig:tira_camp_difforb}
\end{figure}

\begin{figure}[h]
\begin{center}
\includegraphics*[width=6.5cm]{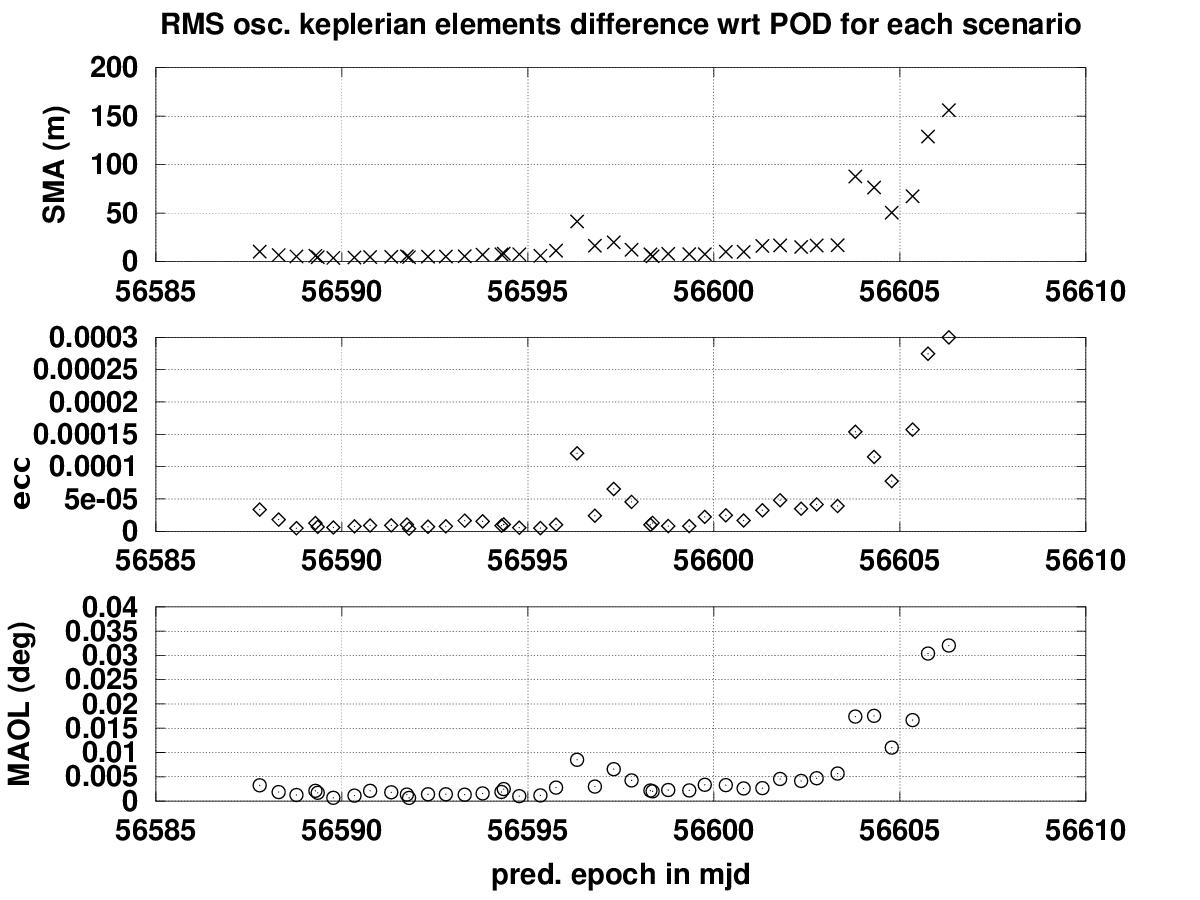}
\end{center}
\caption{RMS of osculating Keplerian elements differences w.r.t. POD
  for each TIRA-based prediction scenario.}\label{fig:tira_camp_diffkep}
\end{figure}

\begin{figure}[h]
\begin{center}
\includegraphics*[width=6.5cm]{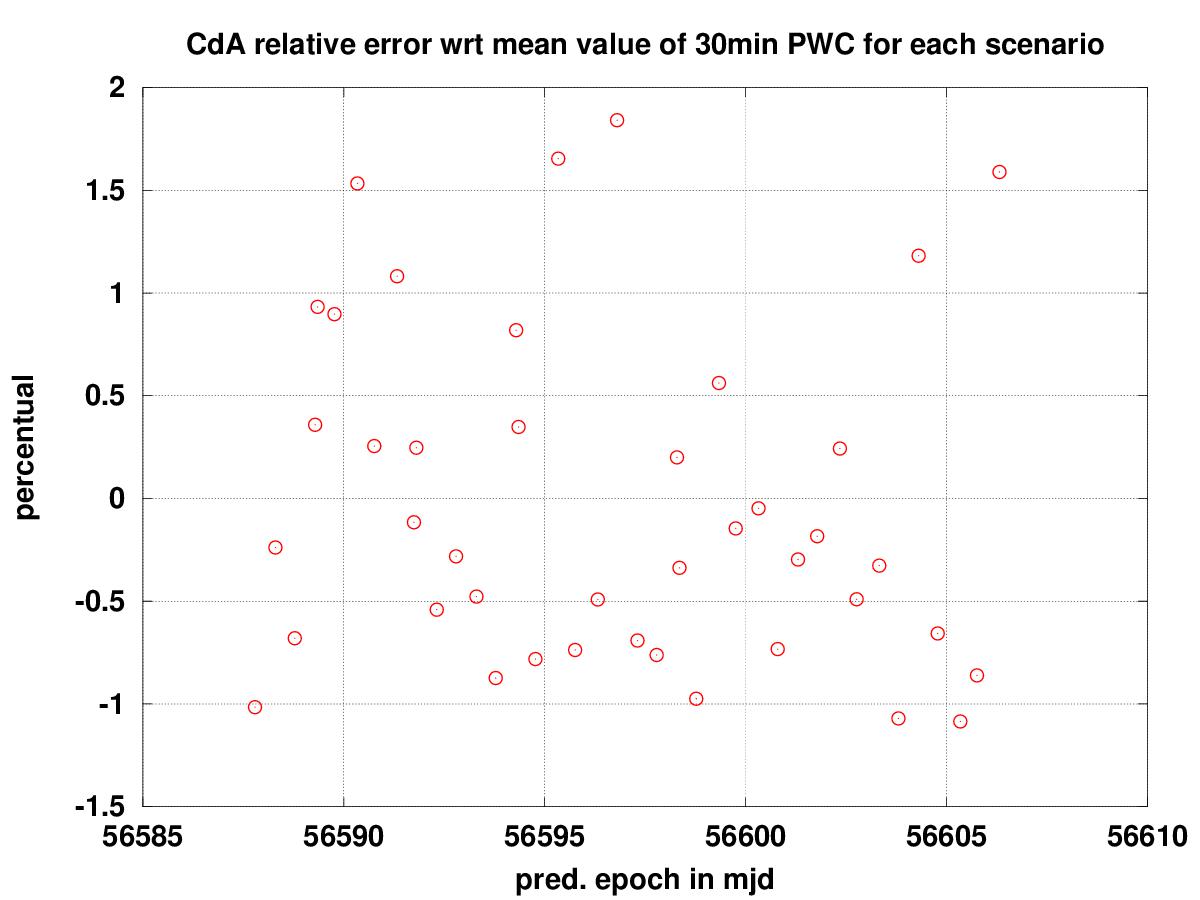}
\includegraphics*[width=6.5cm]{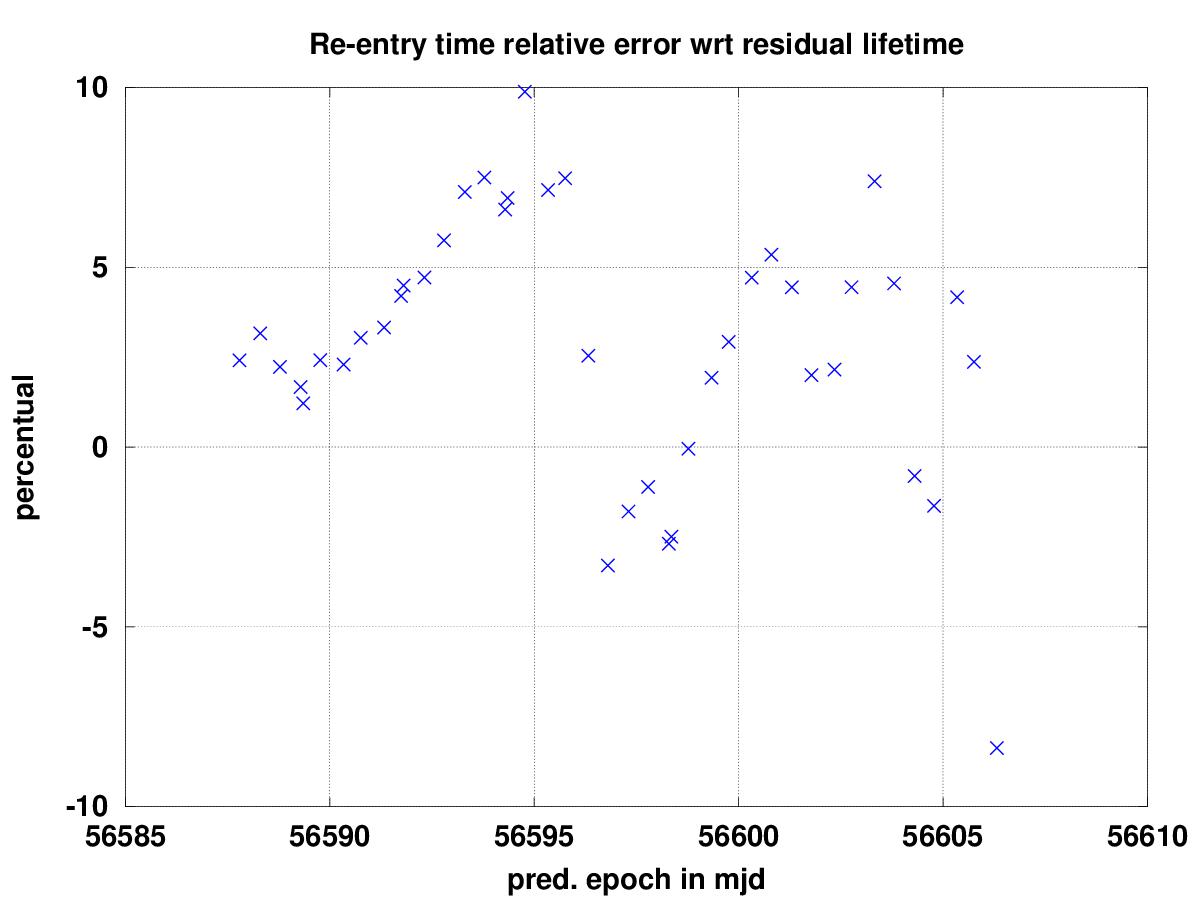}
\end{center}
\caption{lEFT: $C_dA$ coefficient percentual relative difference with
  respect to the mean of the 30min PWC coefficient over each TIRA-based
  prediction scenario's observation time span. Right: Residual
  lifetime percentual relative errors for each prediction of the
  simulated TIRA-based re-entry campaign.}\label{fig:tira_camp_cda}
\end{figure}

We show in Figure~\ref{fig:tira_only_drag} the case of a TIRA-only
simulated re-entry campaign, with ballistic coefficient estimations
over intervals of $\sim $36 hours, compared with the 30min PWC
coefficient from POD. More specific results, in terms of differences
with respect to POD are shown in Figures~\ref{fig:tira_camp_difforb},
\ref{fig:tira_camp_diffkep}, \ref{fig:tira_camp_cda}. Note that for
the evaluation of the error in the re-entry time we have
used the residual lifetime relative error formula:
$(t_{re}-t_{re}^{tr})/(t_{re}^{tr}-t_{pred})\cdot 100$, where $t_{re}$
and $t_{re}^{tr}$ are the estimated and true re-entry epochs,
respectively, and $t_{pred}$ is the epoch of prediction (i.e. the
epoch of the last data available).

\clearpage
In the case of EISCAT UHF radar alone, we have already mentioned the
fact that, since we do not have accurate azimuth/elevation
measurements, and since the duration of tracking is very short, the
data contained in only two or three subsequent passes contain too
little information to obtain a full 7-dim OD solution. We will discuss
these cases in Section~\ref{subsec:crit_cases}. In this section we focus only on
EISCAT-based solutions with four consecutive tracks, which cover about
36h, and in some cases about 48h.

For comparison with Table~\ref{tab:goce_scen1} and
Figure~\ref{fig:tira_cda_estim_sc1}, we show in
Table~\ref{tab:goce_scen1_eis} the results for the OD scenario which
includes the first four EISCAT tracks during re-entry days 1 and 2, in
terms of comparison with the reference orbit, while the estimated
ballistic coefficient in comparison with the POD-based 30min PWC
coefficient has a relative difference of $\sim 1.3\%$. The
corresponding estimated re-entry epoch is Nov-11 at $\sim$12:46 UTC. As we
can notice, the TIRA-based vs. EISCAT-based corresponding results are
very much aligned in terms of re-entry prediction, even though the
orbital differences in the RTW system are significantly different. In
other words, over a comparable total observation time span, the
TIRA-based and EISCAT-based re-entry predictions are equivalent. We
will see that this is true for the entire simulated campaign. We will
come back to this point later in Section~\ref{subsec:orbacc}.
\begin{table}[h]
  \caption{RMS differences between EISCAT-based estimated orbit and POD over re-entry days 1 and 2, in the R-T-W system and in osculating keplerian elements semimajor axis $a$, eccentricity $e$, and mean argument of latitude $\omega+M$}
\begin{tabular}{llllllllll}
\hline
    &$\Delta \mathbf{x}_R$ &$\Delta \mathbf{x}_T$ &$\Delta \mathbf{x}_W$ & $\Delta \mathbf{v}_R$ &$\Delta \mathbf{v}_T$ &$\Delta \mathbf{v}_W$ \\
\hline
RMS & 797.1m &2620.9m &808.6m &2.6m/s &0.9m/s &0.9m/s \\
\hline
  & $\Delta a$ & $\Delta e$ & $\Delta (\omega+M)$ \\
\hline
RMS & 12.2m &$1.2\times 10^{-4}$ & $0.016^\circ$ \\
\hline
\end{tabular}
\label{tab:goce_scen1_eis}
\end{table}

\begin{figure}[h]
\begin{center}
\includegraphics*[width=8cm]{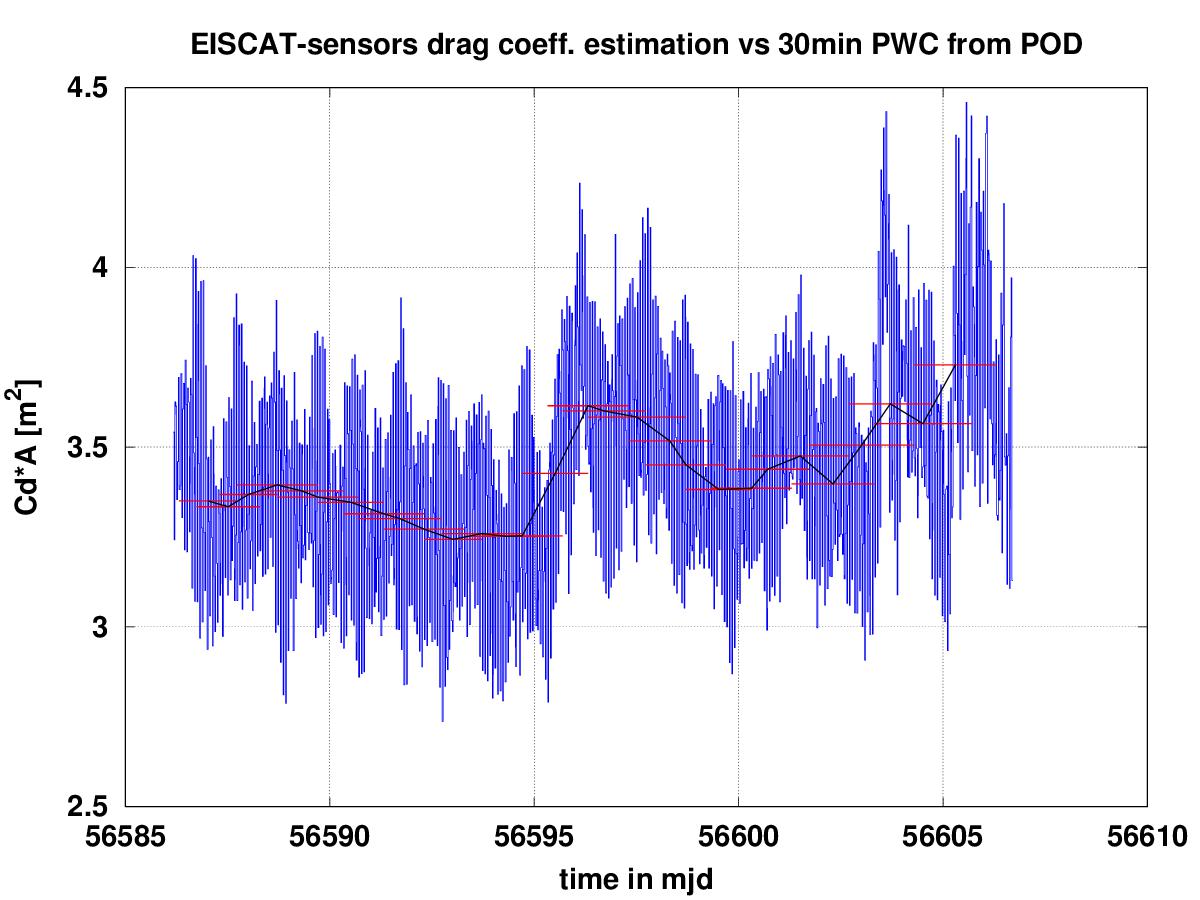}
\end{center}
\caption{EISCAT-only ballistic coefficient estimation over intervals of
  ~36 hours (in red), compared with 30min PWC coefficient from POD (in
  blue).}\label{fig:eiscat_only_drag}
\end{figure}

\begin{figure}[h]
\begin{center}
\includegraphics*[width=6.5cm]{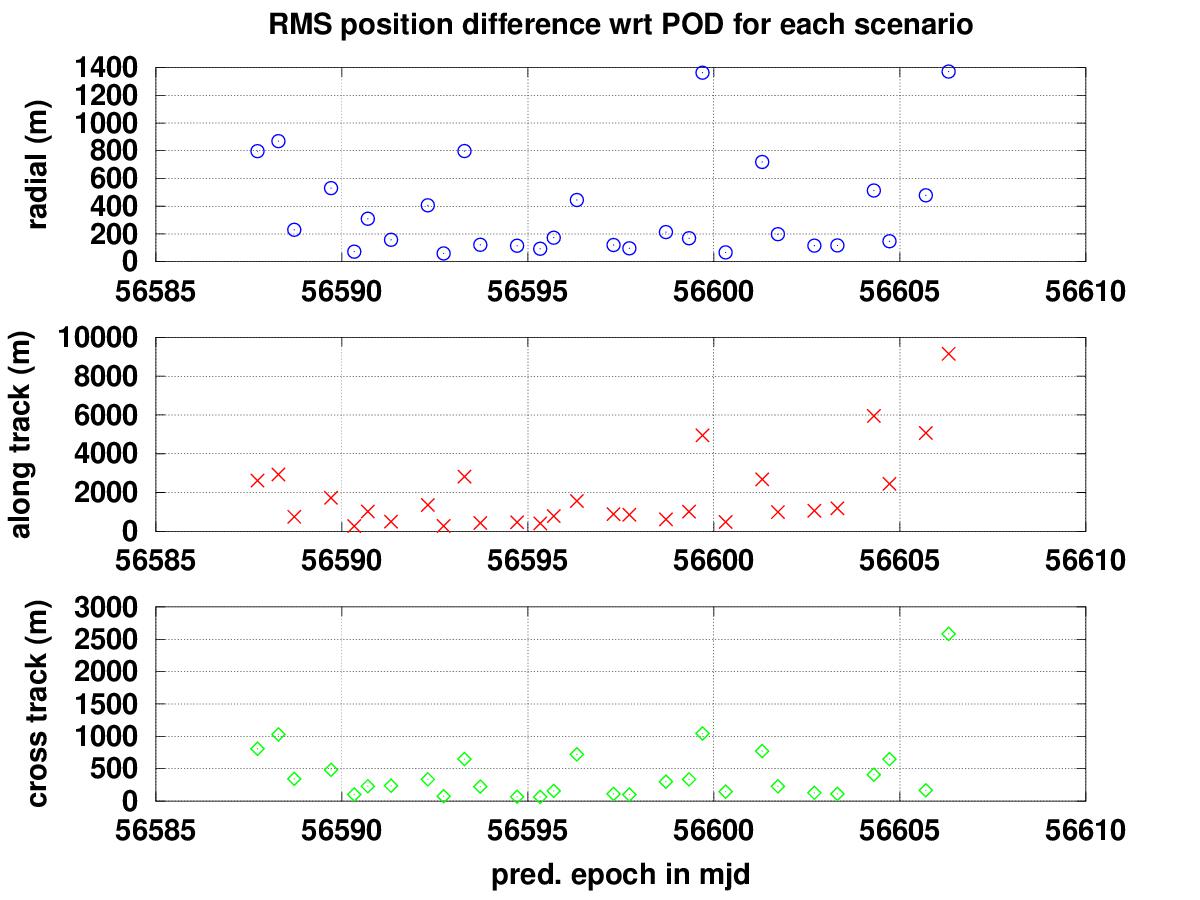}
\includegraphics*[width=6.5cm]{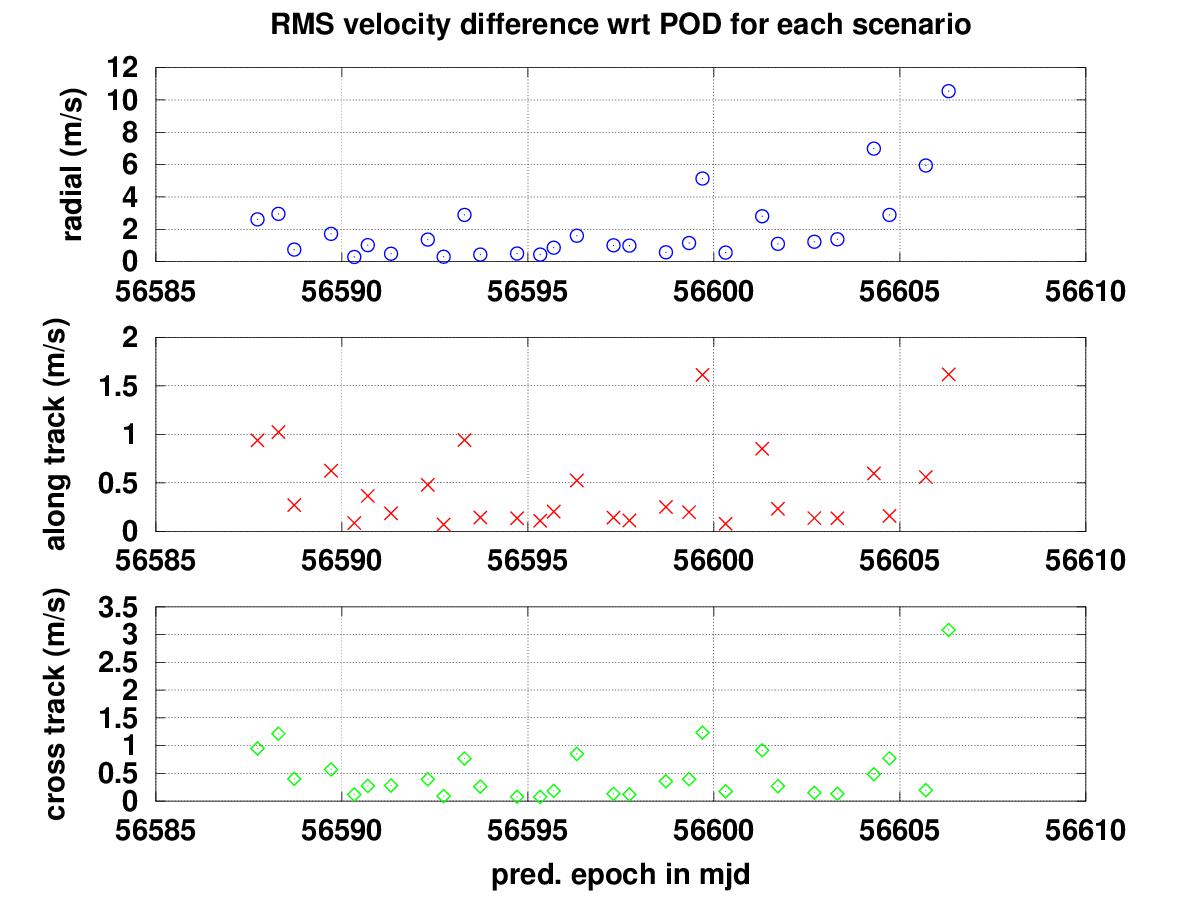}
\end{center}
\caption{RMS of position (left) and velocity (right) differences
  w.r.t. POD for each EISCAT-based prediction scenario.}\label{fig:eiscat_camp_difforb}
\end{figure}

\begin{figure}[h]
\begin{center}
\includegraphics*[width=6.5cm]{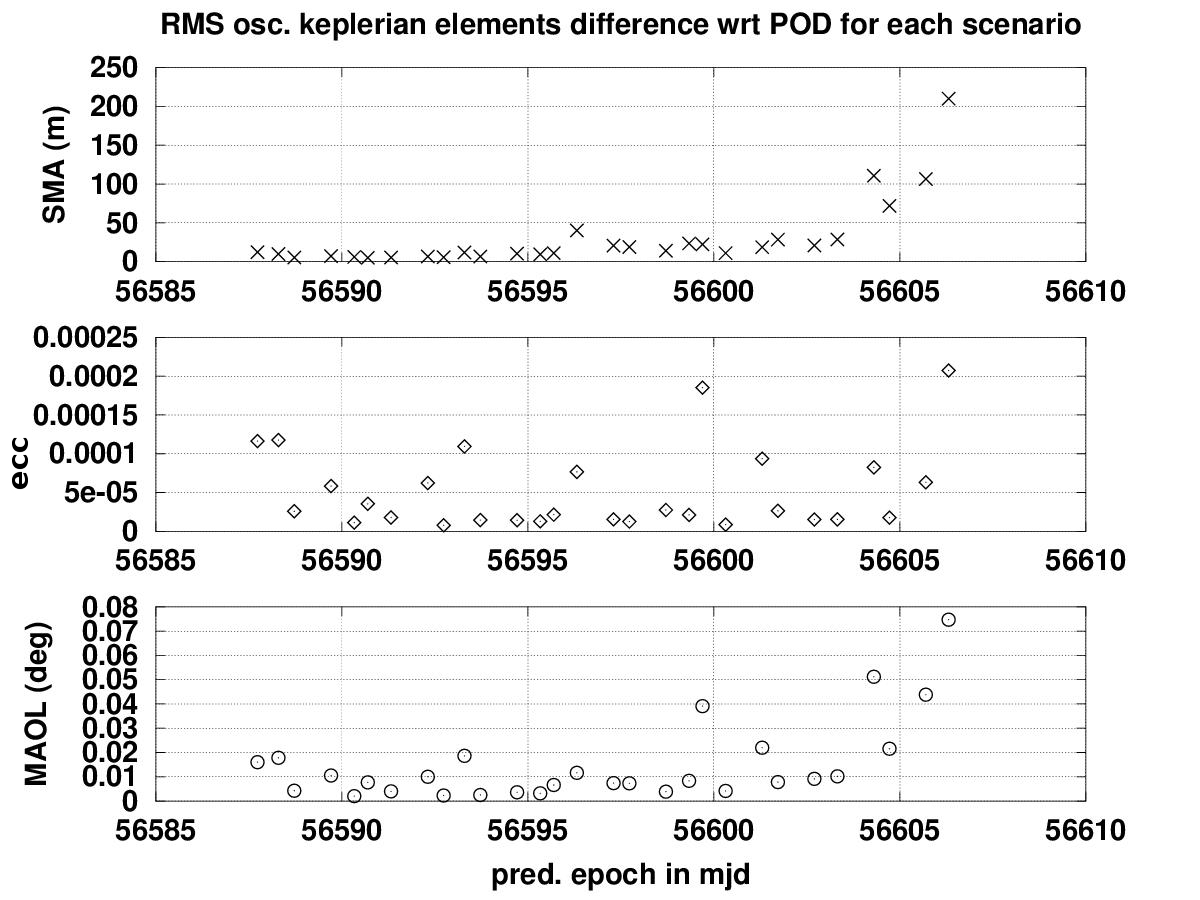}
\end{center}
\caption{RMS of osculating Keplerian elements differences w.r.t. POD
  for each EISCAT-based prediction scenario.}\label{fig:eiscat_camp_diffkep}
\end{figure}

\begin{figure}[h]
\begin{center}
\includegraphics*[width=6.5cm]{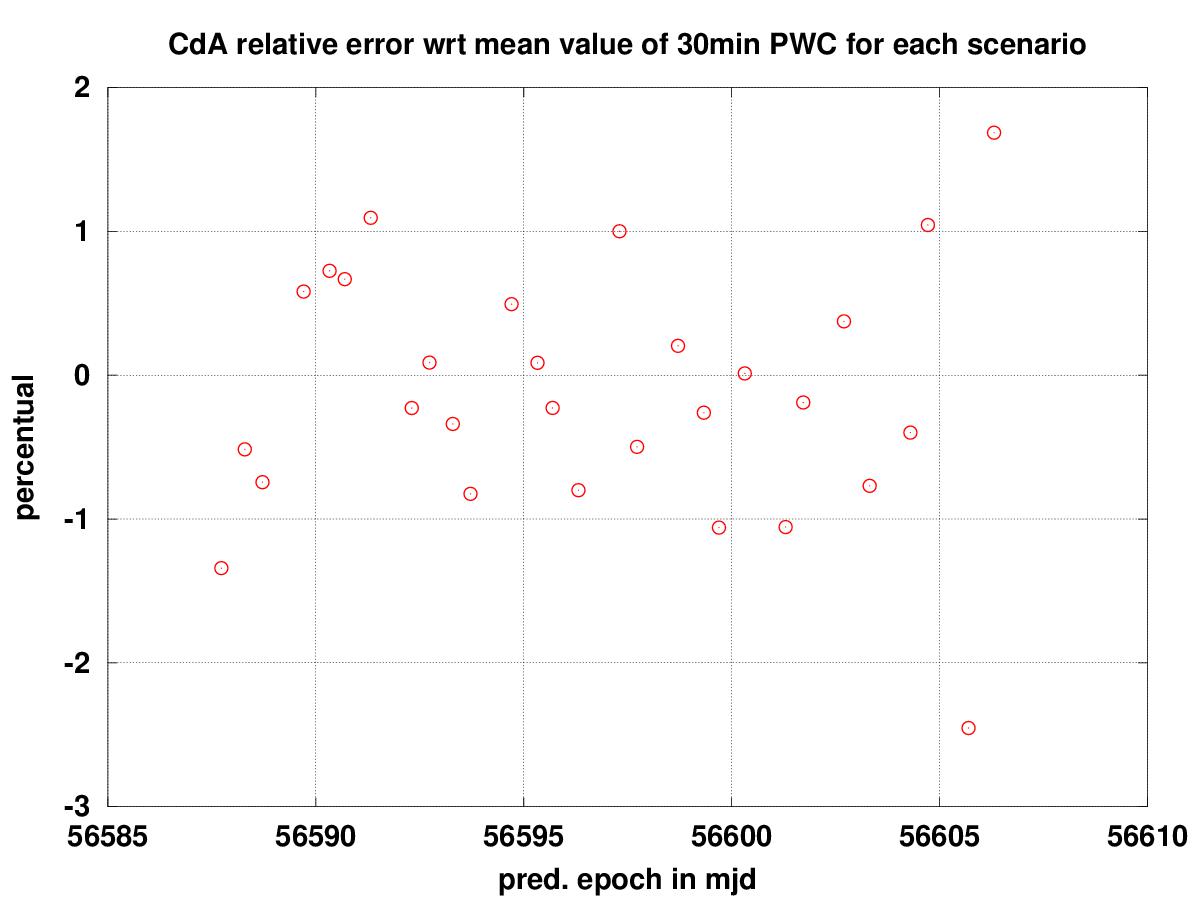}
\includegraphics*[width=6.5cm]{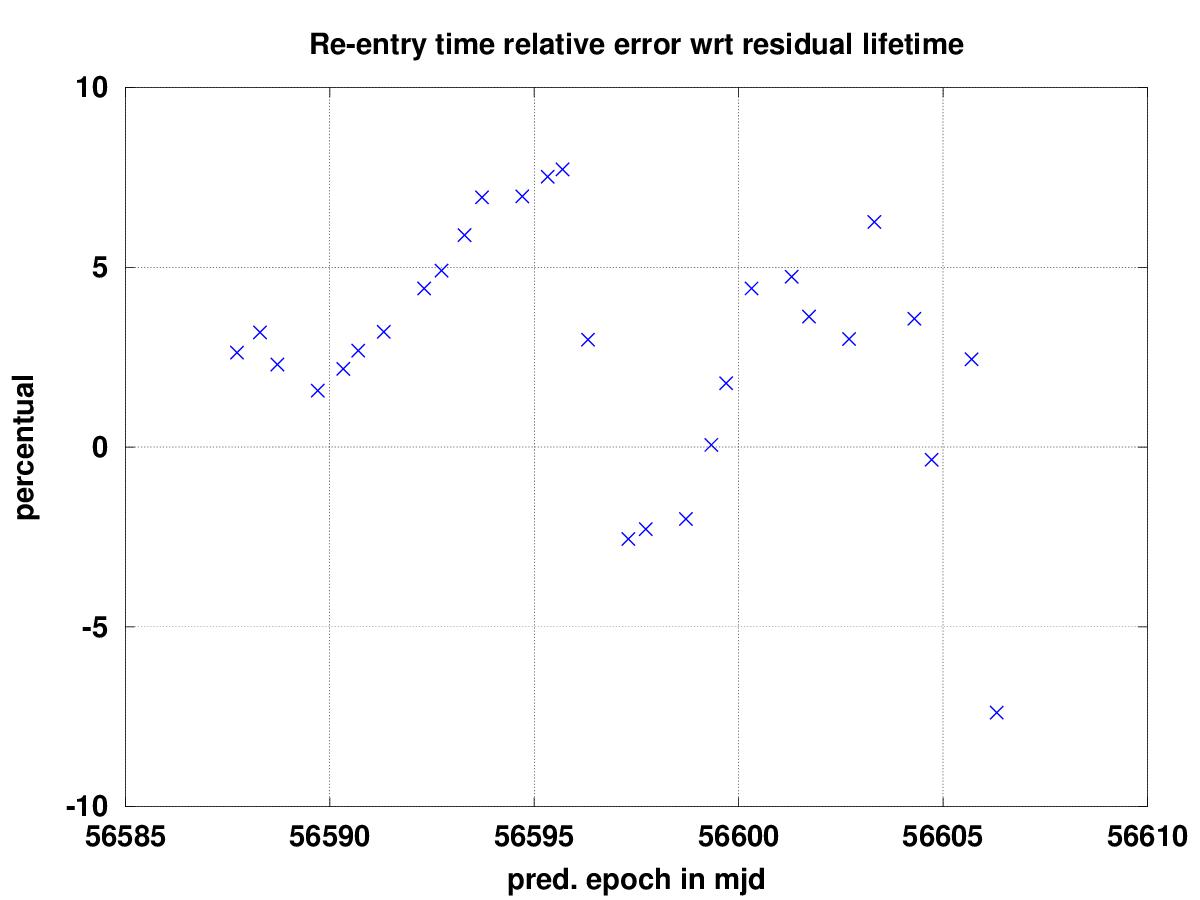}
\end{center}
\caption{Left: $C_dA$ coefficient percentual relative difference with
  respect to the mean of the 30min PWC coefficient over each EISCAT-based
  prediction scenario's observation time span. Right: Residual
  lifetime percentual relative errors for each prediction of the
  simulated EISCAT-based re-entry campaign.}\label{fig:eiscat_camp_cda}
\end{figure}

We show in Figure~\ref{fig:eiscat_only_drag} an EISCAT-only simulated
re-entry campaign, with ballistic coefficient estimations over
intervals of $\sim 36$hours, compared with the 30min PWC coefficient
from POD. More specific results, in terms of differences with respect
to POD are shown in Figures~\ref{fig:eiscat_camp_difforb},
\ref{fig:eiscat_camp_diffkep}, \ref{fig:eiscat_camp_cda}. As we can
see from a comparison with the corresponding TIRA-only results, there
is a very good agreement in the ballistic coefficient calibration and
in the re-entry epoch estimation. Other interesting results is that
while in the first part of decay phase (e.g. first week), where the
dynamical mis-modelings are lower, the TIRA-based orbital solutions
are significantly better than the EISCAT-based ones, meaning that the
observational features still dominates over the dynamical errors. On
the contrary, during the last part of decay (last week), both TIRA and
EISCAT based orbital solutions tend to be of comparable accuracy,
meaning that the dynamical errors dominate over the observational
features. Interestingly, we have seen that in all cases the TIRA-based
and EISCAT-based predictions are analogous. This will be explained in
Section~\ref{subsec:orbacc}.

\clearpage

We can now discuss these results in terms of \emph{(i)} frequency of
re-entry predictions, \emph{(ii)} drag coefficient estimation, and
\emph{(iii)} optimization of resources.
\begin{enumerate}[i]
\item Frequency of re-entry predictions: we have already discussed
  that the TIRA and EISCAT passes are generally close in time
  (Figure~\ref{fig:passes}), due to the quite close longitudes of the
  two stations and the geometrical constraints. Moreover, due to the
  high elevation threshold, the EISCAT sensor can have $\sim $24 hours
  of lack of visibility, thus reducing the frequency of possible
  predictions.
\item Ballistic coefficient estimation: Provided that we can obtain a
  full OD solution with EISCAT data only (at least four tracks), the
  TIRA-based and EISCAT-based ballistic coefficient estimations over
  comparable time spans are very close to each other. On the contrary,
  when we need to estimate the ballistic coefficient over shorter time
  spans (e.g. 12-24h) the EISCAT tracks need to be combined either
  with TIRA measurements or with apriori information (see
  Section~\ref{subsec:crit_cases}).
\item Optimization of resources: As long as we are far from re-entry,
  and ballistic coefficient calibrations over 36-48h are enough to
  have reliable re-entry predictions, then EISCAT and TIRA data can be
  used without significant constraints. On the contrary, when we are
  closer to re-entry, e.g. during the last two days, it is highly
  recommended to have at least one TIRA pass available per day, to
  combine with the EISCAT tracks over shorter time spans.
\end{enumerate}

\clearpage

\subsection{Orbit accuracy and variation of re-entry time}
\label{subsec:orbacc}

In terms of estimation of the re-entry epoch, the EISCAT-only
simulated campaign turns out to be equivalent to the TIRA-only one
over comparable observation time spans. From the point of view of the
overall orbit determination accuracy, the TIRA-based solutions are
better, due to the larger availability of good measurements, at least
in the part of decay where the dynamical mis-modelings are not too
large. In order to understand this fact, we shall focus on the
first week of GOCE decay, which includes about the first ten
predictions of our simulated campaigns.

\subsubsection{Differences with respect to POD}

\begin{figure}[h]
\begin{center}
\includegraphics*[width=6.5cm]{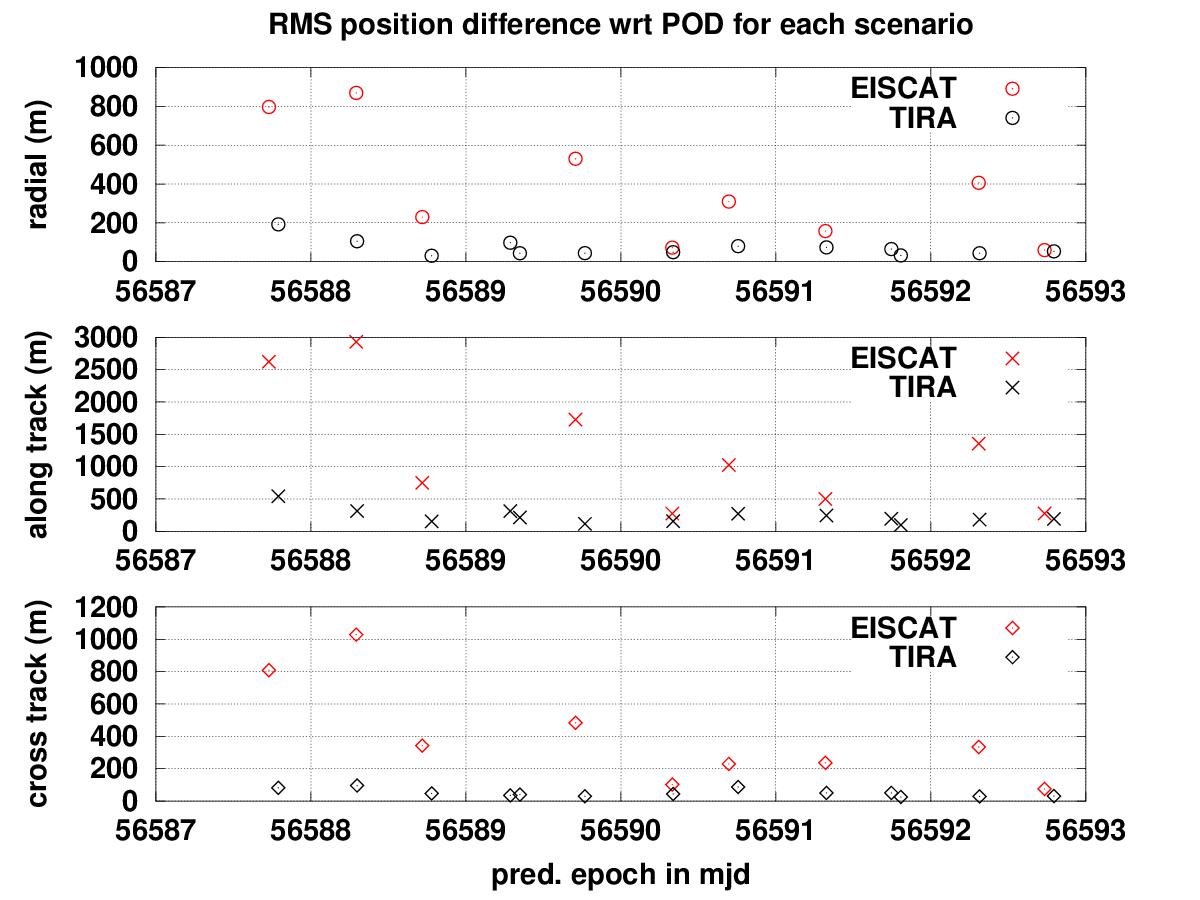}
\includegraphics*[width=6.5cm]{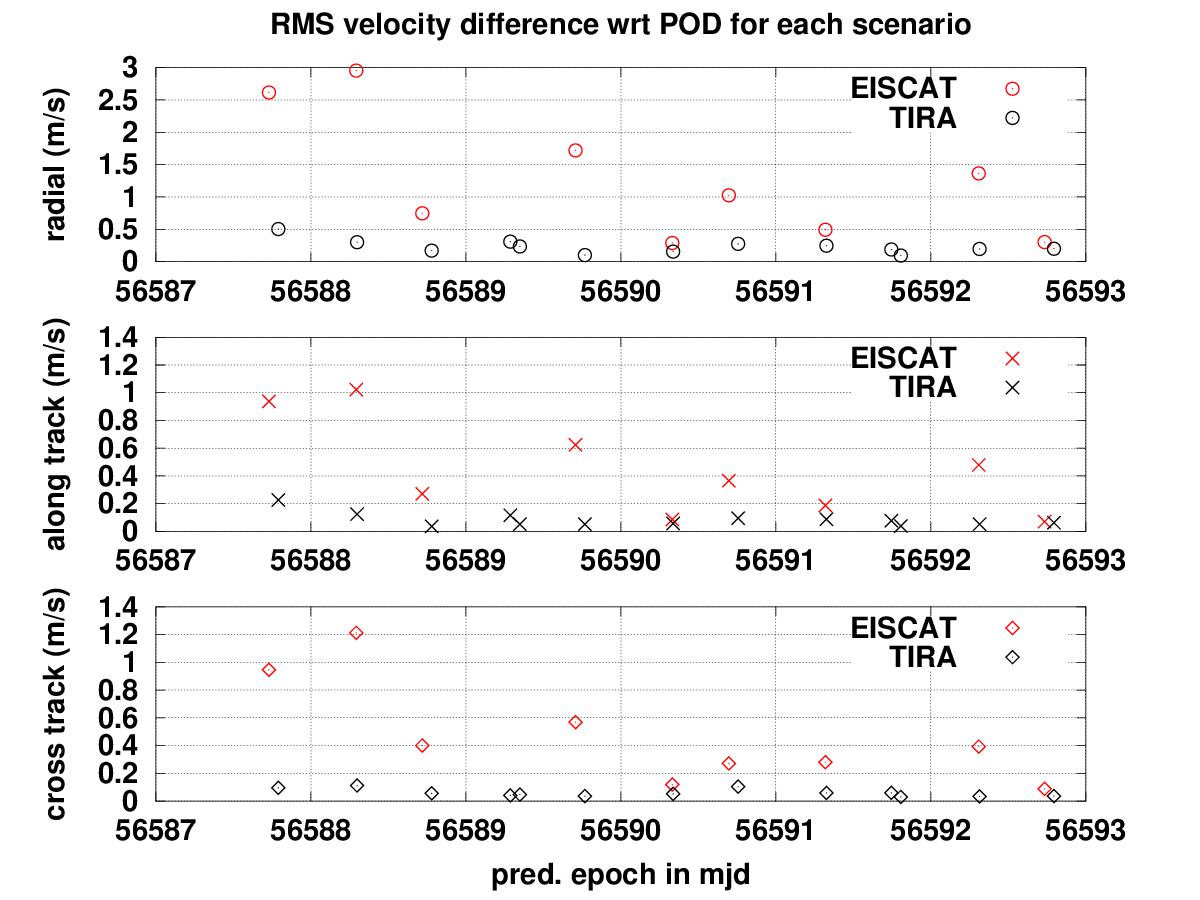}
\end{center}
\caption{RMS of position (left) and velocity (right) differences
  w.r.t. POD for each TIRA and EISCAT-based prediction scenario.}\label{fig:eiscat_tira_camp_difforb}
\end{figure}

We see from Figure~\ref{fig:eiscat_tira_camp_difforb} that the errors
in the RTW reference frame are significantly different for the two
sensors, as an example, the error in radial component can be one order
of magnitude different during the first days of decay. However, it
must be noted that the errors in position and velocity are not
independent and they can show significant correlations. As a matter of
fact, the correlations between the position and velocity errors, over
the time spans of each prediction scenario, can be even larger than
0.99 between radial position and along-track velocity, and along track
position and radial velocity. Since the errors in these quantities are
important for the variations of the corresponding re-entry predicted
time, the RMS of the errors can be an overestimated quantification.
An easier error budget can be obtained from Keplerian elements error
quantification.

\subsubsection{Errors in RTW vs errors in Keplerian}

\begin{figure}[h]
\begin{center}
\includegraphics*[width=6.5cm]{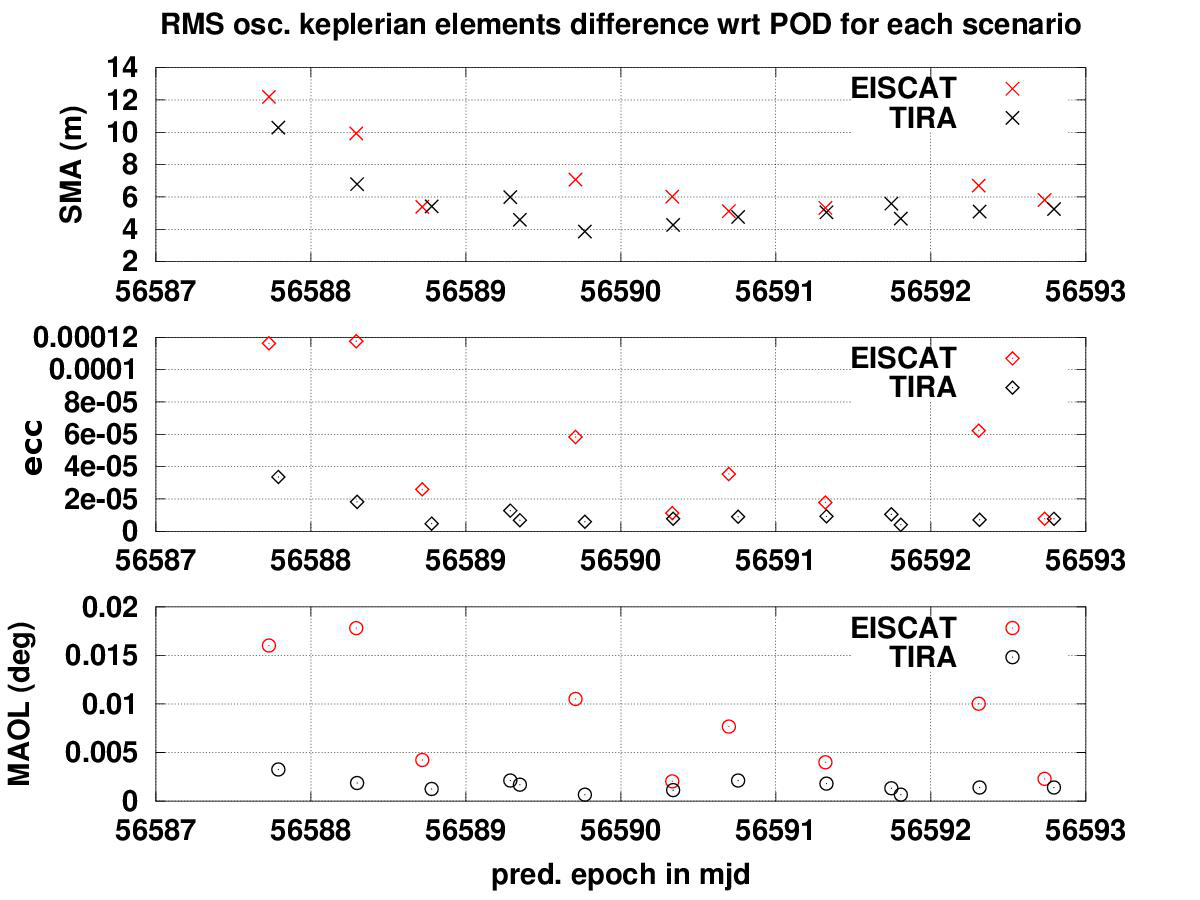}
\includegraphics*[width=6.5cm]{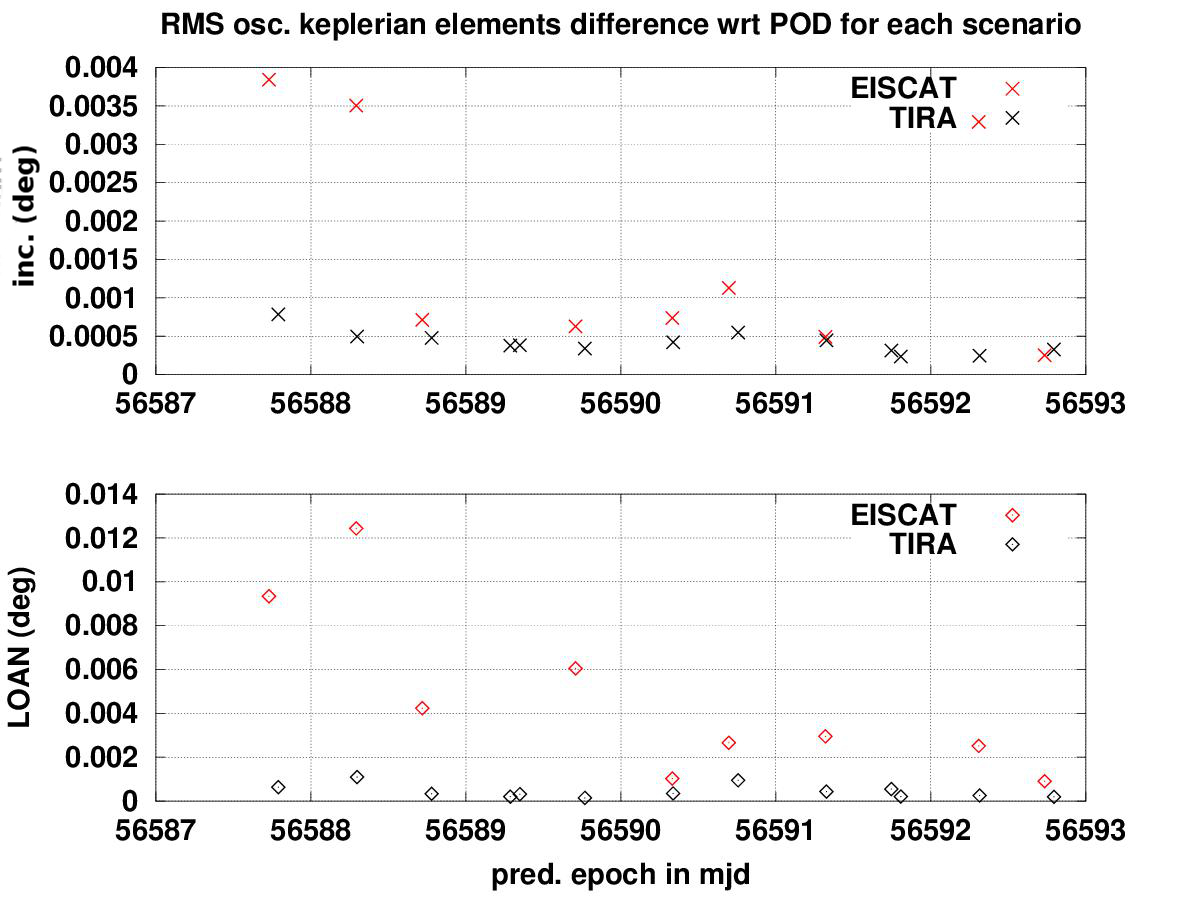}
\end{center}
\caption{RMS of osculating keplerian elements differences w.r.t. POD
  for each TIRA and EISCAT-based prediction scenario (left:
  $a$,$e$,$\omega+M$, right: $i$,$\Omega$).}\label{fig:eiscat_tira_camp_diffkep}
\end{figure}

On the contrary, if we consider the errors with respect to POD in
Keplerian elements (osculating), the larger correlations between the
elements are between argument of perigee $\omega$ and mean anomaly $M$
($>0.99$), but this is due to the low eccentricity of the orbit, and
sometimes between the inclination $i$ and the longitude of the
ascending node $\Omega$. The correlations between semimajor axis $a$,
eccentricity $e$ and mean argument of latitude $\omega +M$ are
generally low ($<0.5$). The comparison between the errors in keplerian
elements with respect to POD for the first prediction scenarios are
shown in Figure~\ref{fig:eiscat_tira_camp_diffkep}. We can notice that
the errors in semimajor axis are very much aligned between the two
sensors. If we want to understand how these input errors affect the
propagated re-entry time we can set up a simplified computation.

\subsubsection{Variation of re-entry time}

We want to set up a simplified computation that gives us a
quantification of the variation of the computed re-entry time in
function of the initial conditions errors, in Keplerian elements. For
this test, we fix the epoch of prediction on Oct-25 (MJD 56590,
Table~\ref{tab:incondtest}), and we let the initial orbital elements
to vary inside an interval. With the nominal initial conditions and a
$C_dA=3.4$m$^2$, the nominal re-entry epoch is on Nov-11 at
7:49:48UTC. The size of the variations for each elements are taken
from Figures~\ref{fig:tira_camp_diffkep},
\ref{fig:eiscat_camp_diffkep} for the corresponding five elements $a$,
$e$, $i$, $\Omega$, $\omega+M$, and increased by a conservative
factor. Then, for each initial condition a re-entry time is computed,
and the final variation can be evaluated.
\begin{table}[h]
  \caption{TLE GOCE osculating keplerian elements on 2013-10-25 at 10:54:45 UTC, in GCRS.}
\begin{tabular}{llllll}
\hline
  $a$  & $e$ & $i$ & $\Omega$ & $\omega$ & $M$ \\
\hline
$6587.553$km & $1.3871 \times 10^{-3}$ & $96.5163^\circ$ & $327.5510^\circ$ & $334.0914^\circ$  & $134.4271^\circ$ \\
\hline
\end{tabular}
\label{tab:incondtest}
\end{table}

\begin{table}[h]
  \caption{Simplified test to check the variation of the re-entry epoch prediction w.r.t. the initial conditions variations.}
\begin{tabular}{lll}
\hline
Initial element & Variation interval & Corresponding variation of residual lifetime  \\
                &                    & (percentual relative error)                   \\
\hline
$a$ & $\pm 300$m & $<1.0\%$ \\
\hline
$e$ & $\pm 3\times 10^{-4}$ & $<0.5\%$ \\
\hline
$\omega +M$ & $\pm 0.1^\circ$ & $<0.5\%$ \\
\hline
$i$ & $\pm 0.01^\circ$ & $<0.01\%$ \\
\hline
$\Omega$ &$\pm 0.05^\circ$ &$<0.01\%$ \\
\hline
\end{tabular}
\label{tab:vartest}
\end{table}

What we can see from the results of Table~\ref{tab:vartest} is that
both TIRA and EISCAT radar-based orbit determination solutions provide
errors that do not change significantly the predicted re-entry
epoch. Similar results hold also for a prediction epoch closer to
re-entry (e.g. on MJD 56605).

\subsubsection{Choice of the calibration time span}

As regards the ballistic coefficient estimation, we can apply a
similar analysis from the results shown in
Figures~\ref{fig:tira_camp_cda} and \ref{fig:eiscat_camp_cda}. In a
RMS sense, the variations of the ballistic coefficient estimation are
of the order of 1-2\% with respect to the nominal value given by the
average of the POD-based 30min PWC coefficient. Analogous computations
show that the variation of the ballistic coefficient affects the
corresponding variation in the predicted re-entry lifetime
approximately by the same percentual value, thus implying that the
errors in the re-entry predictions shown in
Figures~\ref{fig:tira_camp_cda} and \ref{fig:eiscat_camp_cda}, which
are of the order of 10\%, are not mainly due to OD inefficiencies, but
they are dominated by the average mis-modelings occurring in the time
span from the prediction epoch to the actual re-entry. For example,
they can be due to unpredicted significant attitude changes and
unmodeled atmospheric density variations. Strictly speaking, we can
say that these re-entry predictions are quite precise but not very
accurate (low trueness).

We can see from Figures~\ref{fig:tira_only_drag} and
\ref{fig:eiscat_only_drag} (and similars in the next
Sections~\ref{sec:tle_calib} and \ref{sec:add_sim}) that the choice of
the observation time span is crucial to reconstruct the main
variations of the ballistic coefficient. As a matter of fact, for a
fixed prediction epoch, different calibration time spans can lead to
different ballistic coefficients (they capture the average over a
different time span) and thus to different re-entry predictions. At
this point, the problem is not to understand if one calibration gives
a better prediction than the other, because we know that the
correctness of the prediction mostly depend on what happens after the
prediction epoch. The main problem is to understand, from what we know
from the calibrations up to the current one, what is the general
attitude behaviour of the object (e.g. stable, tumbling) and try, if
possible, to predict if major changes can occur afterwards. The same
applies to the prediction of significant space weather events, and
the atmospheric environment variations. However, this can be quite
difficult.

The results for the cases analysed in this study indicate that, during
the last weeks of decay, intervals of $\sim$36-48h are good enough to
reconstruct the average behavior of the ballistic coefficient
variations. In general, it is recommended to perform ballistic
coefficient calibrations over different time spans, when this is
possible, provided that we have enough data avaliable. 

In the next Section~\ref{sec:tle_calib} we will perform the same kind
of OD and ballistic coefficient calibration analysis by using TLE data
only.

\subsection{Critical cases}
\label{subsec:crit_cases}

We have pointed out many times that there are some critical cases for
which the standard OD and ballistic calibration does not work
properly. First of all, if we have less than four very short tracks
from EISCAT, of range and range-rate, the solutions are in general bad
determined (or even ill-posed). Second, if we are performing OD close
to re-entry, even if we have enough observational data to compute a
full solution and good initial conditions, in some cases the dynamical
systematic errors are particularly strong to introduce instabilities
in the differential corrections and even divergence. This problem can
occur for TIRA data processing as well, for instance when we try to
fit all the last five TIRA passes together (time span $\sim$48h). We
discuss in the following some techniques that can help when dealing
with these cases.

\

\textbf{1. Only two EISCAT passes.} When only two EISCAT passes are
available, such as for example the last two tracks shown in
Figure~\ref{fig:passes}, a good option would be to ask for the
additional availability of a TIRA tracking pass, for example the last
one. The significant amount of information contained in the TIRA pass,
combined with the two EISCAT pass and a good initial condition
(e.g. TLE-based), will lead to a more stable problem and possibly to a
good OD and ballistic coefficient estimation.  If an additional TIRA
pass is not available, but at least the initial condition has a
reliable error estimation, then imposing an apriori constraint on the
initial position and velocity will lead to a more stable problem and
possibly to a good OD and ballistic coefficient estimation. For
example, in the cases considered in our tests, if the initial
condition is a TLE at an epoch close to the observations times, an
empirically derived $\sim$1km constraint in position and $\sim$1m/s in velocity
turned out to be a reasonable and effective option.

\

\textbf{2. Only three EISCAT passes.} The case with three EISCAT
passes is quite similar to the previous one, and it can be treated in
the same way.  However, so far we have considered only range and
range-rate short tracks, without assuming any information available
for the azimuth and elevation angles of the tracked object. If
confirmed, a quite low accuracy information could be deduced from the
radar pointing direction, for example to the $1-2^\circ$ level. Adding this
information on the direction of the tracked object would help in
stabilizing the problem and possibly lead to a good OD and ballistic
coefficient estimation.

\

\textbf{3. Unstable differential corrections.} We have always pointed
out that, when we approach the re-entry, the altitude decreases and
the mis-modelings in the non gravitational perturbations grow in
magnitude causing large errors in the estimated orbits. In general we
have seen that, given a proper observational scenario, it is possible
to find a good solution for re-entry predictions anyway, at the cost
of obtaining large residuals with respect to observational noise.

In some cases, even if we have enough observational data, and good
initial conditions, it is difficult to compute a full OD and ballistic
coefficient estimation because the problem shows instabilities, and
the differential corrections diverge. This is the analogous of
``over-shooting'' effects in the framework of Newton's method for
finding the roots of a generic non linear function.  We believe that
this problem is mainly due to a combined effect of intrinsic, even
small, weaknesses in the OD covariance matrix, and particularly large
systematic dynamical errors that affect the residuals. In other words,
an analogous observational scenario in presence of lower systematic
errors would firmly converge to a good solution. A typical example is the
data processing of the last five TIRA observational passes, or the
last four EISCAT tracks, which cover $\sim$48h from 2013-Nov-9 to
2013-Nov-10.

We have tried at least three experimental strategies that could help
in leading the problem to converge, or at least to approach to a good
solution:
\begin{itemize}
\item damped differential corrections (under-relaxation), 
\item differential corrections with pseudoinversion (descoping), 
\item use of a-priori constraints on initial conditions with deweighed observations.
\end{itemize}
The first one is recommended when we believe that the observations are
enough to have a well-posed OD problem, while the other two are more
useful when we have more intrinsic weaknesses in the OD problem.  We
now briefly describe how these strategies can be applied, to give an
idea on what are the main formulas involved. However, this description
is not intended to be a guarantee of success in finding always a good
solution, and may need a more rigorous treatise to be generalized.

If we generically indicate with $\mathbf{x}$ the 7-dimensional vector
of solve for parameters (6-dim state vector + 1 ballistic
coefficient), with $B_x$ the partial derivatives of the measurements
residuals $\mathbf{\xi}$ with respect to $\mathbf{x}$, and with W the
weight matrix of the observations, then the normal equations to be
solved during the batch least squares differential corrections at
iteration $k$ are: $C\Delta \mathbf{x} = \mathbf{D}$, where $\Delta
\mathbf{x}=\mathbf{x}_k-\mathbf{x}_{k-1}$, $C=B_x^TWB_x$ is the normal
matrix, and $\mathbf{D}=-B_x^TW\xi$. The solution is found by
inverting the normal matrix $\Gamma = C^{-1}$, and computing $\Delta
\mathbf{x} = \Gamma \mathbf{D}$. More specifically, the inverse matrix
can be found by the diaginalization of $C$ with a orthonormal base of
eigenvectors: $C=U^TEU$, $\Gamma = UE^{-1}U^T$, where $U$ is an
orthonormal matrix having the eigenvectors of $C$ as columns, and $E$
is a diagonal matrix having the eigenvalues of $C$ as elements.
According to \cite[Chap.6]{mg10}, given a proper choice of the units
of measure, if the confidence ellipsoid defined by $C$ is very much
elongated along a particular direction, this direction is given by the
eigenvector corresponding to the smallest eigenvalue and is called
weak direction. Since, during the differential correction steps, the
component of $\Delta \mathbf{x}$ along each eigenvector is multiplied
by the inverse of the corresponding eigenvalue, it may happen that the
correction along the weak direction is so large to keep the state
$\mathbf{x}$ far from the convergence basin.

The main idea here is to force the process to avoid large differential
corrections, and check if this succeeds in finding a minimum of the
least squares target function. Then such a solution is compared with
the reference orbit as usual, to see if it is good enough for a
re-entry prediction.

\

\textbf{3.1 Damped differential corrections.} In this case the idea is
to apply a damping factor to the entire correction vector: $\Delta
\mathbf{x}=\Gamma \mathbf{D} / \alpha$, where $\alpha = MAX(1,n-k+1)$, $k$ is
the current iteration, and $n$ is the number of damped
corrections. This process can be iterated, i.e. after $n$ damped
corrections we can restart from the current state and compute $n$ new
damped corrections. This technique is a type of under-relaxation
method for the differential corrections iterative process, with
$1/\alpha$ is the under-relaxation factor.

\

\textbf{3.2 Differential corrections with pseudoinversion.} If
$\lambda_j$ are the eigenvalues of $C$, in increasing order of
magnitude, and $\mathbf{v}_j$ are the corresponding eigenvectors, then
the differential correction can be also written as $\Delta \mathbf{x}
= \sum_j \lambda_j^{-1} (\mathbf{v}_j\cdot \mathbf{D})\mathbf{v}_j$. In
order to avoid large corrections along the weak direction
$\mathbf{v}_1$, the pseudoinverse technique consists in relpacing
$\lambda_1^{-1}=0$ in the $E^{-1}$ matrix, thus $\Delta \mathbf{x} =
\sum_{j>1} \lambda_j^{-1} (\mathbf{v}_j\cdot \mathbf{D})\mathbf{v}_j$, and
the correction is performed in the hyperplane orthogonal to
$\mathbf{v}_1$ (see also \cite[Chap.10]{mg10}). The same strategy can be
applied on more eigenvalues/eigenvectors, restricting the corrections
to lower dimensional hyperplanes.

\

\textbf{3.3 A priori constraints on initial conditions with deweighed
  observations.} Another differential correction strategy which could
help in finding a good solution in a critical case consists in
exploiting an apriori contraint on the initial conditions, such as the
one introduced for the cases of only two or three EISCAT
tracks. However, what happens here is that the formal covariance
matrix $\Gamma$ has already small diagonal terms, because the problem
is due to the systematic dynamical errors. To let the apriori
constraint help in finding a solution, it is possible to try a uniform
deweigh of the observations, i.e. $W \rightarrow W/\beta^2$, where
$\beta$ is a suitable tuning parameter. One possibility is to choose
$\beta$ in order to have the diagonal terms of $\Gamma$ at the same
order of magnitude of the a priori constraint variances.


\section{TLE-based ballistic coefficient calibration}
\label{sec:tle_calib}

Exploiting TLE information to calibrate re-entry predictions is a
common approach. Adopting the same stretegy described in the previous
Section~\ref{sec:gocesim}, we could use all the TLE available during the three weeks
of the GOCE re-entry to perform a piece-wise ballistic coefficient
calibration to be compared with the POD-based one.  

\begin{figure}[h]
\begin{center}
\includegraphics*[width=8cm]{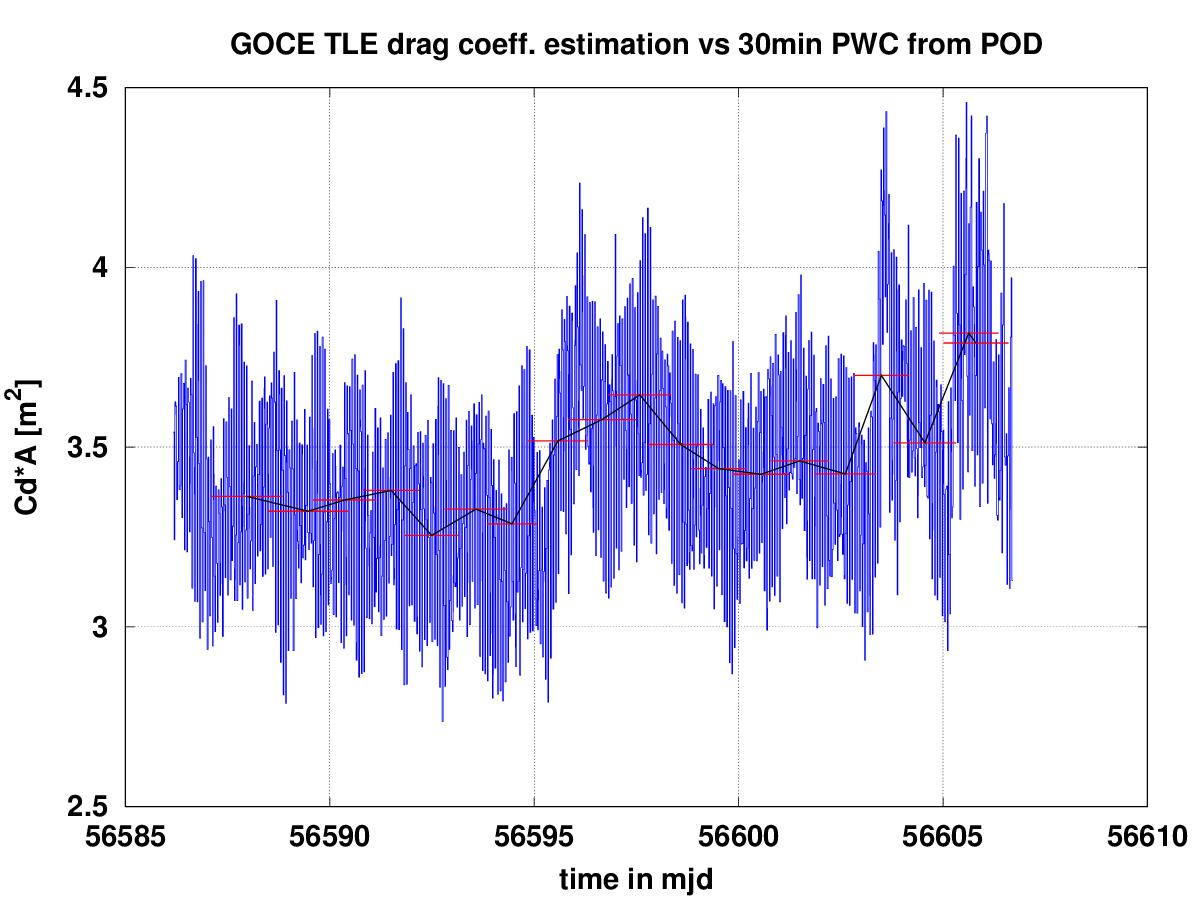}
\end{center}
\caption{GOCE TLE-based ballistic coefficient estimation over intervals of
  ~36 hours (in red), compared with 30min PWC coefficient from POD (in
  blue).}\label{fig:goce_tle_drag}
\end{figure}

The procedure simply consists in taking all the TLE available over
time intervals of 1-2 days, and perform a full OD and ballistic
coefficient estimation by using the TLEs as state observations
(i.e. TLE-fitting). The weights to be applied to each TLE's position
and velocity in the fit would need a general knowledge of the
accuracy/covariance of its corresponding estimation. Such information
is not officially available, and in these experiments we have applied
a general data weighing of 1km in position and 1m/s in velocity. The
results are shown in Figure~\ref{fig:goce_tle_drag}. As described in
Figures~\ref{fig:tira_only_drag}, \ref{fig:tira_camp_cda} for the TIRA
calibrations, and Figures~\ref{fig:eiscat_only_drag},
\ref{fig:eiscat_camp_cda} for the EISCAT calibrations, also in the
TLE-based case the average value of the 30min PWC ballistic
coefficient is reconstructed quite well, consistently below the 3\%
level (computed over each calibration time span).

\

\textbf{2012-006K AVUM R/B}

\

\begin{figure}[h]
\begin{center}
\includegraphics*[width=8cm]{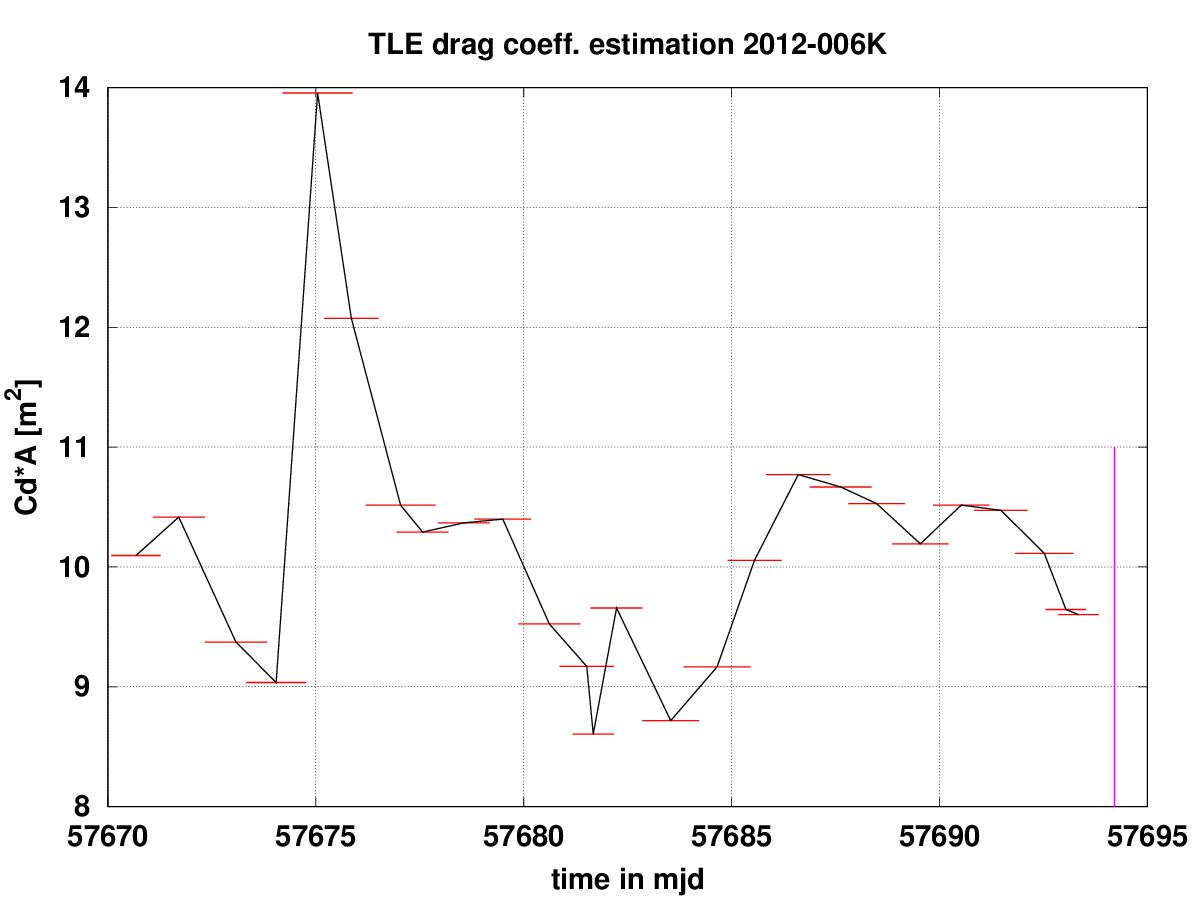}
\end{center}
\caption{2012-006K TLE-based ballistic coefficient estimation (in
  red), over intervals of $\sim$36h. In magenta the nominal re-entry
  epoch.}\label{fig:tle_drag_avum}
\end{figure}

\begin{figure}[h]
\begin{center}
\includegraphics*[width=6.5cm]{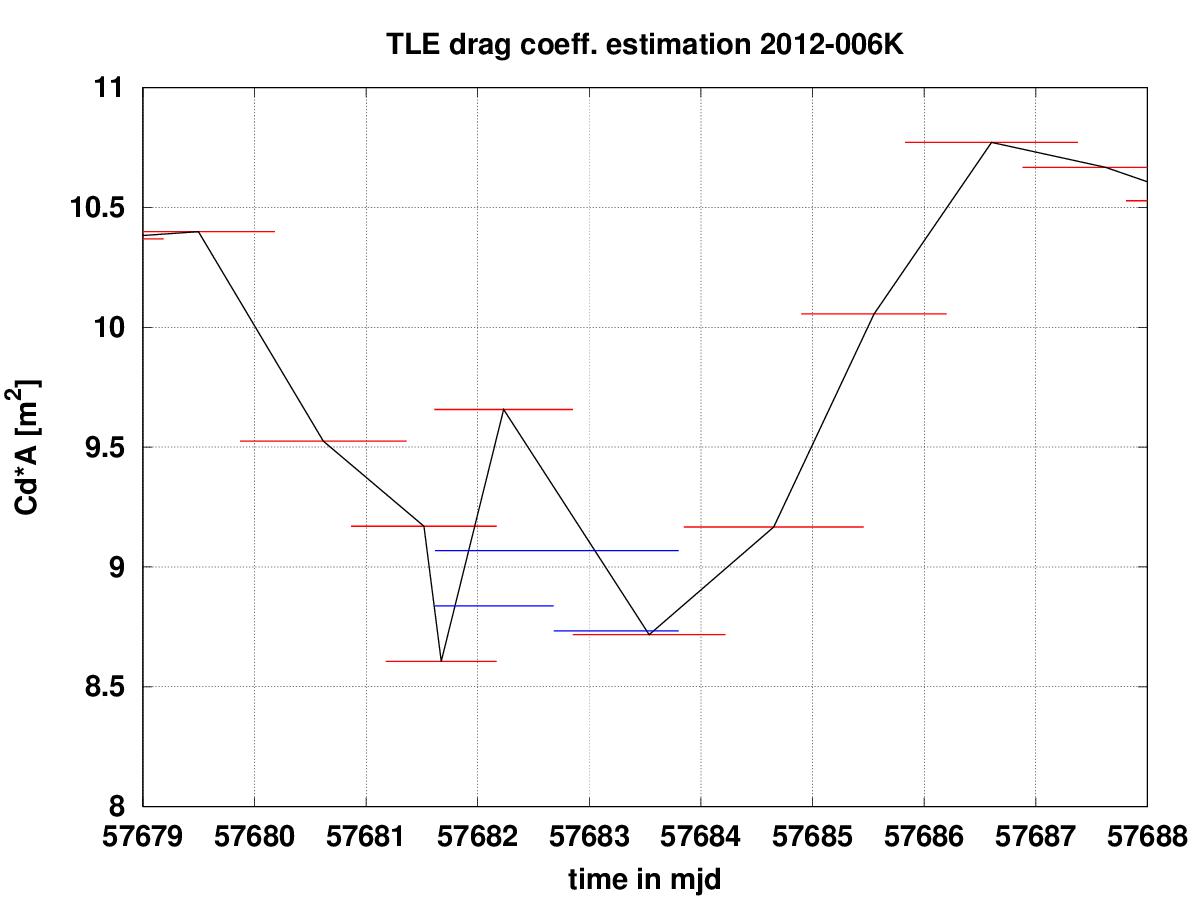}
\includegraphics*[width=6.5cm]{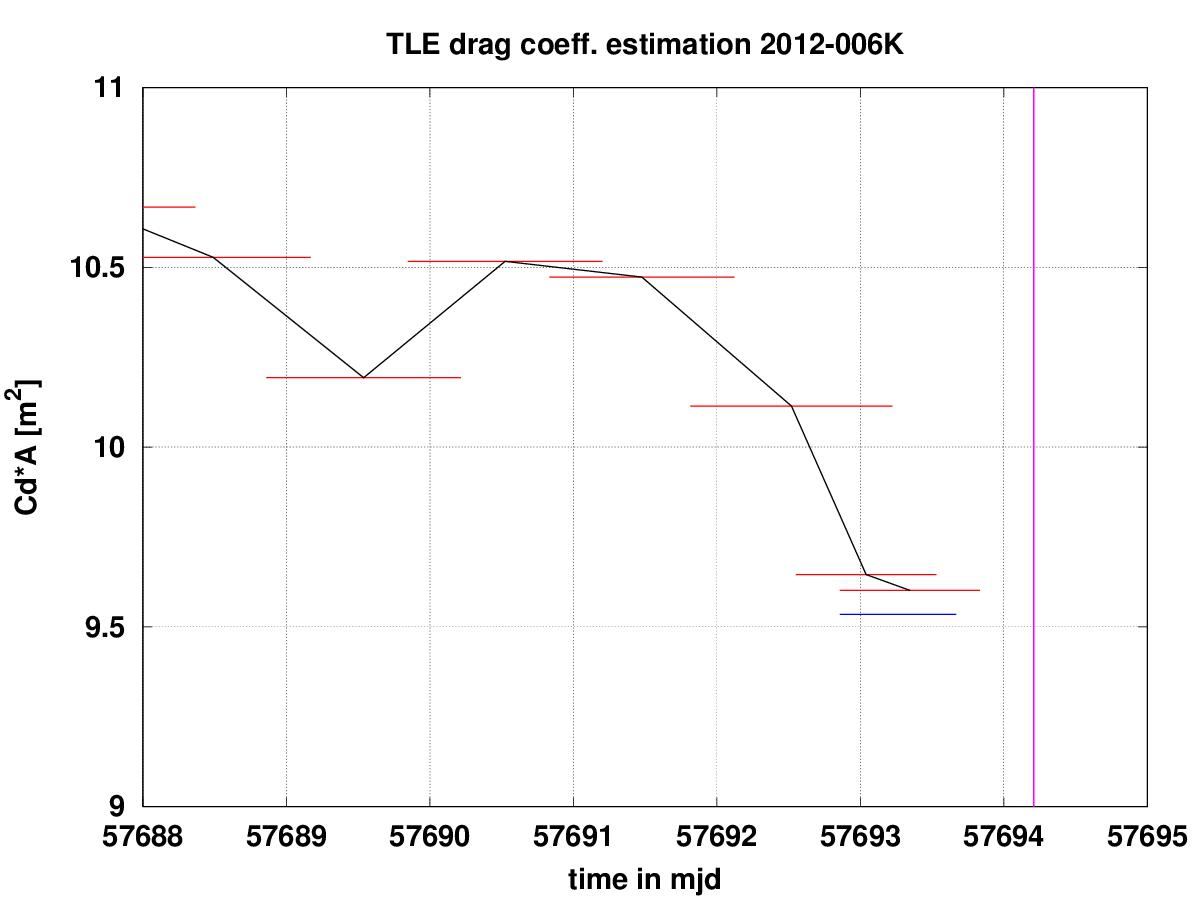}
\end{center}
\caption{2012-006K TLE-based ballistic coefficient estimation (in
  red), over intervals of $\sim$36h. Left: In blue, the radar-based
  calibrations described in Table~\ref{tab:avum_res}. Right: In blue,
  the radar-based calibration described in
  Table~\ref{tab:avum_scen4_res}.}\label{fig:tle_rad_drag_avum}
\end{figure}

The same experiment can be performed for the 2012-006K AVUM rocket
body, and it can be compared with the results computed with the real
data given in Section~\ref{sec:2012-006K}. The results are shown in
Figures~\ref{fig:tle_drag_avum} and \ref{fig:tle_rad_drag_avum}.

As we can recognize in Figure~\ref{fig:tle_rad_drag_avum}, the
TIRA/EISCAT-based calibration are in good agreement with the general
behavior of the ballistic coefficient variations around their epoch of
prediction.

\clearpage

\section{Additional simulations with different types of objects}
\label{sec:add_sim}

In order to do a first step in the generalization of the previous
analysis, we want to perform EISCAT-based ballistic coefficient
calibration also for the simulated objects discussed in
\cite{cetal17}, which are a spherical and a cylindrical object.

\subsection{Sphere}
\label{subsec:sphere}

\begin{figure}[h]
\begin{center}
\includegraphics*[width=8cm]{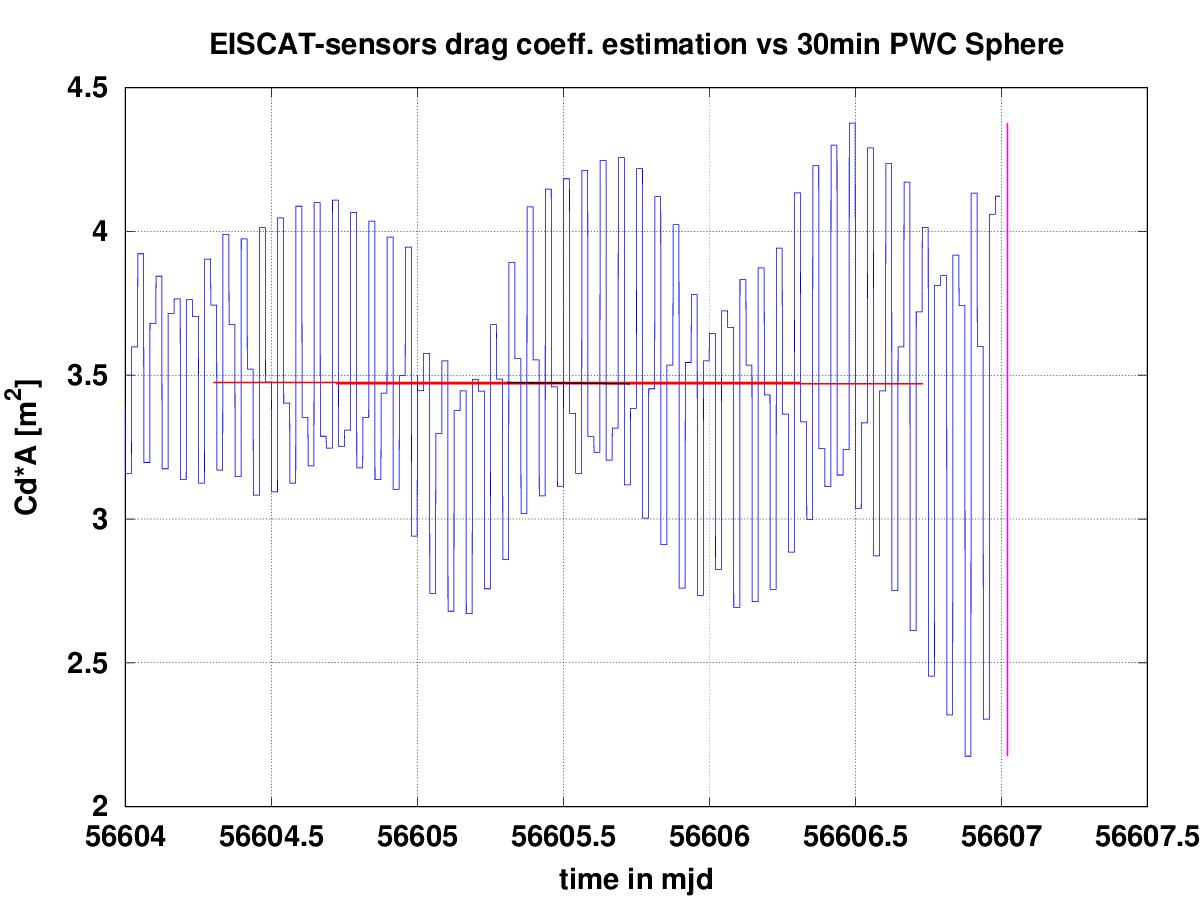}
\end{center}
\caption{EISCAT-based ballistic coefficient calibration of a
  sphere-like object (in red), compared with 30min PWC coefficient based on
  reference orbit (in blue).}\label{fig:sphere_eiscat_only_camp}
\end{figure}

The first object we consider is one with a constant ballistic
coefficient. We generated the orbital motion of a sphere-like object
that stays close to GOCE quite enough, in order to obtain the same
visibility conditions (as in Figure~\ref{fig:passes}). This simulated
object has a re-entry at 90km on NOV-11$\sim$00:45 UTC. The useful
aspect of the spherical object is that we do not have uncertainties in
its cross sectional area or attitude motion, hence we can test the
effects due to different atmospheric models. To this purpose, we
simulated the reference orbit with a Jacchia-Bowman 2008 density
model, and to test the behaviour of a 30min PWC drag coefficient
estimation with the NRLMSISE00 density model. As regards the OD tests,
also in this case we have considered sets of four subsequent EISCAT
tracks. We already stressed that the radar-based orbit determination
always tries to capture the average of this function, and this is
confirmed also in this case, see
Figure~\ref{fig:sphere_eiscat_only_camp}. Due to the lack of attitude
variations, in this simplified test the average value of the ballistic
coefficient is very stable, even if large short-term oscillations due
to simulated atmospheric density errors are present.

\newpage
\subsection{Tumbling cylinder}
\label{subsec:tumbcyl}

\begin{figure}[h]
\begin{center}
\includegraphics*[width=8cm]{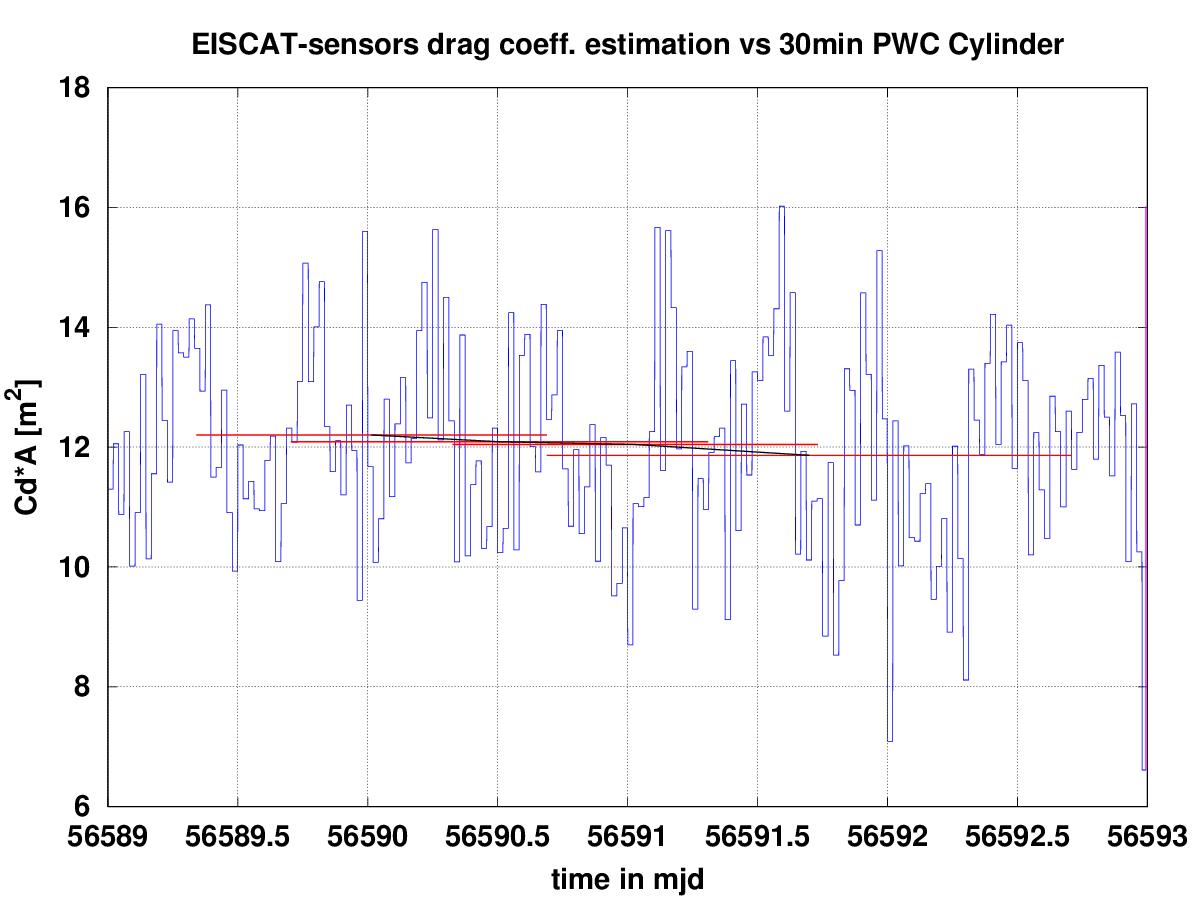}
\end{center}
\caption{EISCAT-based ballistic coefficient calibration of a tumbling
  cylinder object (in red), compared with 30min PWC coefficient based on
  reference orbit (in blue).}\label{fig:cyl3_eiscat_only_camp}
\end{figure}

The tumbling cylinder object is an essentially unoriented cylinder, it
has a small centre of gravity offset of $-0.03$m in order to establish
tumbling motion within the re-entry period, whose orbit was generated
by Belstead Research Ltd. 6DOF propagator\footnote{SpaceDyS's parter
  during the project (\url{belstead.com/ats6.html}).} (see also
\cite{cetal17}). It has the same size of GOCE, and it is propagated
from the same initial conditions of GOCE on OCT-22.  It has a re-entry
at 90km on OCT-27$\sim$23:52UTC. The atmospheric density model used in
the simulator for the orbit generation is a dynamic
Jacchia-Roberts. In this case, the tumbling increases the average
cross-section of the object exposed to drag, thus the absolute
magnitude of the ballistic coefficient increases. However, the overall
tumbling motion turned out to have a quite stable average value, and
we can see in Figure~\ref{fig:cyl3_eiscat_only_camp} that this is well
captured by the radar-based calibrations.

\newpage
\subsection{Tumbling cylinder with final stabilization}
\label{subsec:tumbcyl_stab}

Finally, the third object that we analyse is an oriented-unstable
cylinder, it is a cylinder with a forward centre of gravity of $+0.32$m,
which will give it a favoured orientation, but starting from an
initial unstable, backward orientation. The result is a vehicle which
tumbles, but aligns in the final two days. It is propagated from the
same initial conditions of GOCE on OCT-22. The tumbling motion
significantly raises the average drag leading to a shorter trajectory,
and it has indeed a re-entry at 90km on OCT-28$\sim$1:50UTC.

\begin{figure}[h]
\begin{center}
\includegraphics*[width=8cm]{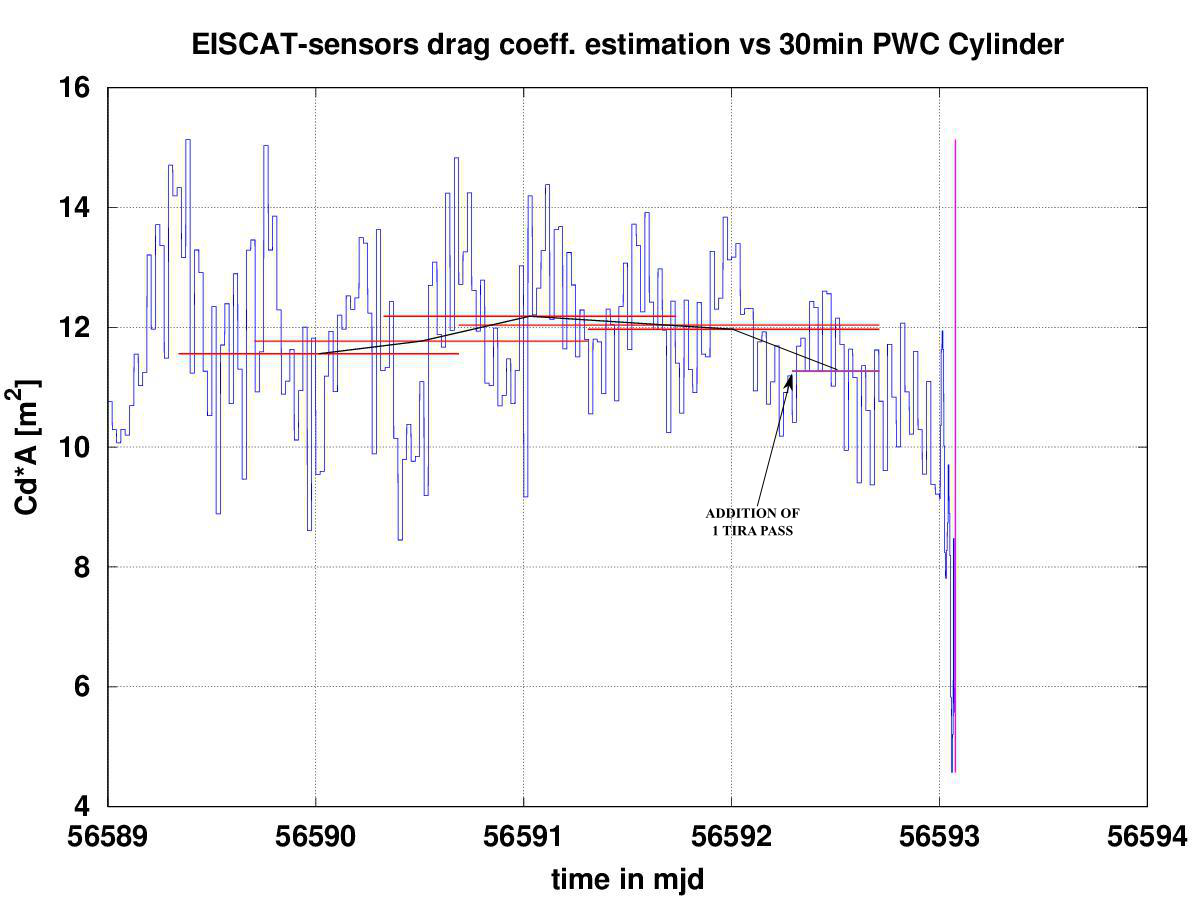}
\end{center}
\caption{EISCAT-based ballistic coefficient calibration of a tumbling
  cylinder object with final attitude stabilization (in red), compared with
  30min PWC coefficient based on reference orbit (in blue). In the last day of
  decay an additional TIRA pass is used for a short-time calibration.}\label{fig:cyl2_eiscat_only_camp_wtira}
\end{figure}

Since the object tends to align in the final two days, a significant
descrease in the long-term average value is expected in the last part
of decay. This can be noted in
Figure~\ref{fig:cyl2_eiscat_only_camp_wtira}, where the EISCAT-based
calibrations are effective for the most part of the decay. However, if
we are forced to use only four EISCAT pass per calibration,
unfortunately in this case we are not able to detect the last
$\sim$48h decrease in the ballistic coefficient average value. As
discussed before, the addition of just one TIRA pass, if available,
during the last day of decay can allow for a shorter-term
calibration.

\clearpage

\section{Conclusions}
\label{sec:concl}

During this work we have carried out additional analysis on
radar-based re-entry prediction calibrations for GOCE, and for other
similar decaying objects on circular and highly inclinated
orbits. After the analysis described in \cite{cetal17}, where the main focus
was on the german TIRA radar, we focused here on the northern european
sensor EISCAT UHF radar, located in Troms\o, Norway. This sensor,
originally conceived for atmospheric studies of the ionosphere, has
been recently considered for space debris applications, and in
particular for tracking of specific targets, and to support re-entry
predictions. The limited tracking capabilities of this sensor posed
the problem of establishing to which extent it would be useful to
support orbit determination and re-entry predictions, in comparison to
what we know about TIRA-like standard prediction calibrations.

By exploiting the large amount of information coming from the re-entry
of GOCE, for instance the GPS-based POD and a refined piecewise
constant estimation of its ballistic coefficient variations, during
the three weeks of uncontrolled decay, we have set up a realistic
simulation scenario for a re-entry campaign. Radar-based re-entry
predictions and ballistic coefficient calibrations were performed and
compared in the cases of TIRA-only, EISCAT-only, and both radar
availability situations. The results were compared in terms of
differences in the orbital states over the total observation time
span, differences in the ballistic coefficient estimation, and in the
corresponding re-entry epoch.  The main conclusion is that, provided a
minimum amount of necessary observational information, EISCAT-based
re-entry predictions are of comparable accuracy to TIRA-based (but
also to GPS-based, and TLE-based if TLE errors are properly accounted)
corresponding ones. Even if the worse tracking capabilities of the
EISCAT sensor are not able to determine an orbit at the same level of
accuracy of the TIRA radar, it turned out that the estimated orbits
are anyway equivalent in terms of re-entry predictions, if we consider
the relevant parameters involved and their effects on the re-entry
time. What happens to be very important is the difficulty in
predicting both atmospheric and attitude significant variations in
between the current epoch of observation and the actual re-entry. For
completeness, a corresponding GOCE ballistic coefficient estimation
based on TLE only was presented, and it turned out to be very
effective as well.  Some critical cases which consist in a minimum
amount of observational information, or in difficulties in obtaining
OD convergence, were presented, and a list of possible countermeasures
was proposed.

An experiment with real EISCAT (and TIRA) measurements was also
presented, for the case of 2012-006K AVUM R/B, which re-entered on
2016-Nov-2. A corresponding TLE-based ballistic coefficient
calibration was performed for comparison. The results turned out to be
reliable and compatible with each other.  Finally, some numerical
experiments with simulated trajectories were presented, for the cases
of a spherical object, and a cylindrical tumbling object. In both
cases we have confirmed the same kind of conclusions obtained for the
GOCE case, with an effective estimation of the average behaviour of
the ballistic coefficient during the decay phase, with EISCAT data
only and EISCAT plus TIRA data available.

In conclusion, from an orbit determination point of view, provided a
suitable and reasonable minimum amount of observational data, and of
corresponding accuracy information, it is possible to compute reliable
and quite precise re-entry predictions. On the contrary, from a more
general point of view, the absolute accuracy of each prediction is
very much affected by the difficulties in predicting future
atmospheric enviroment variations, and significant attitude
changes. For this reason, it is not easy to keep the actual accuracy
of the predictions much lower than 10\% of the residual lifetime,
apart from particular cases with constant area to mass ratio, and low
atmospheric environment variations with respect to current models.

Future activities on this topic could include at least analogous
analysis for objects in more eccentric orbits, and/or with different
shapes and attitude behaviours.


\section*{Acknowledgments}

This work was carried out under ESA Contract No. 4000115172/15/F/MOS
``Benchmarking re-entry prediction uncertainties''. The support of the
ESA Space Debris Office is gratefully acknowledged.


\end{document}